\documentclass[twocolumn,secnumarabic,amssymb, nobibnotes, aps, prd]{revtex4-1}
\pdfoutput=1 
\usepackage[T1]{fontenc} 
\usepackage[utf8]{inputenc}
\usepackage{amsfonts}   
\usepackage{graphics}
\usepackage{epsfig}    
\usepackage[english]{babel}   
\usepackage{caption}  
\usepackage{subcaption} 
\usepackage{amsmath}  
\usepackage{bm} 
\usepackage{multirow} 
\usepackage{array,booktabs} 
\usepackage{calc} 
\usepackage[table]{xcolor}
\usepackage{color}   
\usepackage{makecell} 

\definecolor{aqua}{rgb}{0.0, 1.0, 1.0}
\definecolor{babyblue}{rgb}{0.54, 0.81, 0.94}
\definecolor{beaublue}{rgb}{0.74, 0.83, 0.9}
\definecolor{blizzardblue}{rgb}{0.93, 0.93, 0.93} 
\definecolor{cyan}{rgb}{0.0, 1.0, 1.0}
\newcolumntype{?}{!{\vrule width 0.8pt}} 

\def\tanb{\tan\beta}
\def\hpm{H^\pm}
\def\mhpm{m_{\hpm}}  
\def\mh{m_h} 
\def\mH{m_H}  
\def\ma{m_A}
\def\tb{\tan\beta}

\def\bma{\beta-\alpha}  
 
\newcommand{\RN}[1]{ 
  \textup{\uppercase\expandafter{\romannumeral#1}}%
\renewcommand\thesubfigure{(\alph{subfigure})} 
\captionsetup[sub]{labelformat=simple} 
}
 
\begin{document}

\title{\boldmath Search for neutral Higgs bosons within Type-I 2HDM \\ at future linear colliders} 
\author{Majid Hashemi }
\email{majid.hashemi@cern.ch} 
\author{Gholamhossein Haghighat}
\email{hosseinhaqiqat$@$gmail.com}  
\affiliation{Physics Department, College of Sciences, Shiraz University, \\ Shiraz, 71946-84795, Iran} 
 
\begin{abstract}
In this study, we investigate observability of the neutral scalar ($H$) and pseudoscalar ($A$) Higgs bosons in the framework of the Type-\RN{1} 2HDM at SM-like scenario at a linear collider operating at $\sqrt s=$ 500 and 1000 GeV. The signal process chain $e^- e^+ \rightarrow A H \rightarrow ZHH\rightarrow jj b\bar{b}b\bar{b}$ where $jj$ is a di-jet resulting from the $Z$ boson decay and $b\bar{b}$ is a $b$ quark pair, is assumed and several benchmark scenarios with different mass hypotheses are studied. The assumed signal process is mainly motivated by the possible enhancements the decay modes $A\rightarrow ZH$ and $H\rightarrow b\bar{b}$ may receive in the Type-\RN{1}. Event generation is performed for the assumed scenarios separately and the beamstrahlung effects are taken into account. The detector response is simulated based on the SiD detector at the ILC and the simulated events are analyzed to obtain candidate mass distributions of the Higgs bosons. According to the results, the top quark pair production process has the most contribution to the total background and is, however, well-controlled. Results indicate that, in all of the considered scenarios, both of the Higgs bosons $H$ and $A$ are observable with signals exceeding $5\sigma$ with possibility of mass measurement. To be specific, at $\sqrt s=500$ GeV, the region of parameter space with $m_H=150$ GeV and $200\leq m_A \leq 250$ GeV is observable at the integrated luminosity of 500 $fb^{-1}$. Also, at $\sqrt s=1000$ GeV, the region with $150\leq m_H \leq 250$ GeV and $200\leq m_A \leq 330$ GeV with a mass splitting of 50-100 GeV between the $H$ and $A$ Higgs bosons is observable at the same integrated luminosity.
\end{abstract}

\maketitle 
\flushbottom
  
\section*{Introduction}   
The Standard Model (SM) of elementary particles has emerged as a significant achievement explaining plenty of phenomena. The existence of the Higgs boson, as one of the most striking predictions of the SM, attracted much attention even before its experimental verification \cite{HiggsObservationCMS,HiggsObservationATLAS}. The prediction of the Higgs boson was a direct consequence of the assumed scalar structure which was chosen to be the simplest possible one. To be specific, the SM assumes a single $SU(2)$ doublet with four degrees of freedom as the Higgs field leading to the prediction of a single Higgs boson \cite{Englert1,Higgs1,Higgs2,Kibble1,Higgs3,Kibble2}. However, there are some important motivations, namely the SM inability to explain the universe baryon asymmetry \cite{Trodden}, supersymmetry \cite{MSSM1}, axion models \cite{KIM1}, etc., which provide possibility of an extended scalar structure which leads to the prediction of existence of additional Higgs bosons. Assuming a scalar structure based on two $SU(2)$ Higgs doublets, as one of the simplest possibilities, leads to the Two-Higgs-Doublet model (2HDM) \cite{2hdm_TheoryPheno,2hdm1,2hdm2,2hdm3,2hdm4_CompositeHiggs,2hdm_HiggsSector1,2hdm_HiggsSector2,Campos:2017dgc} which provides interesting environment and phenomenological features. The two assumed Higgs doublets in the 2HDM carry eight degrees of freedom, three of which are absorbed by the three of the electroweak gauge bosons and the remaining five degrees of freedom finally lead to the prediction of five Higgs bosons. The lightest Higgs boson ($h$) is assumed to be the same as the discovered SM Higgs boson (the SM-like scenario) and the four additional Higgs bosons are thought of as yet undiscovered particles. These Higgs bosons include a neutral scalar ($H$), a neutral pseudoscalar ($A$) and two charged ($H^\pm$) Higgs bosons. The present paper concentrates on the $H$ and $A$ Higgs bosons and investigates observability of these particles in several benchmark points in the parameter space of the 2HDM at SM-like scenario.
 
The 2HDM, in its general formulation, predicts tree level flavor-changing neutral currents (FCNCs) which is not in agreement with experimental observations. However, imposing the discrete $Z_2$ symmetry, the tree level FCNCs are well avoided and four 2HDM types with different Higgs-fermion coupling scenarios which naturally conserve flavor are obtained. Different types with different coupling scenarios provide different interesting characteristics \cite{2hdm_HiggsSector2}. The Type-X 2HDM was studied earlier to investigate the observability of the $H$ Higgs boson with the help of large enhancements the Higgs-lepton coupling provides at high $\tanb$ values and led to promising results \cite{H-2HDMX}. In this study, the Type-\RN{1} 2HDM is chosen as the theoretical framework and the signal process through which the Higgs bosons observability is investigated is assumed to be $e^- e^+ \rightarrow A H\rightarrow ZHH \rightarrow jj b\bar{b}b\bar{b}$ where $jj$ is a di-jet resulting from the $Z$ boson decay and $b\bar{b}$ is a $b$ quark pair resulting from the scalar Higgs $H$ decay. The pseudoscalar Higgs boson $A$ undergoes the decay mode $A\rightarrow ZH$ which is dominant due to the SM-like assumption and also the non-zero mass splitting assumed between the $A$ and $H$ Higgs bosons. Scenarios with equal Higgs masses were also considered earlier leading to promising results \cite{MHashemiMMahdavi-1-4b}. The assumed signal process is also motivated by the large enhancement the scalar Higgs decay mode $H\rightarrow b\bar{b}$ receives in low $\tanb$ regime. Such an enhancement is due to the $H$-fermion coupling in the Type-\RN{1} which depends on $\cot\beta$. Moreover, the $A$-$Z$-$H$ vertex depends on $\sin(\bma)$ which is set to unity because of the SM-like assumption. Hence, the process $e^- e^+ \rightarrow A H\rightarrow ZHH$ is independent of $\tanb$ and thus, the scalar Higgs decay mode $H\rightarrow b\bar{b}$ may receive large enhancements at low $\tanb$ values without any destructive effect on the $ZHH$ production process. 
 
Several benchmark points with different mass hypotheses are assumed and investigation of the Higgs bosons observability is done separately for each scenario. Masses of the Higgs bosons are chosen from intermediate and heavy mass regions and thus, the center-of-mass energy of 1000 GeV provides observation possibility in a relatively large portion of the space parameter. However, the center-of-mass energy of 500 GeV is also studied in this study since this energy is easily accessible to future linear colliders and it will be shown that a few scenarios can also be observed at this center-of-mass energy. The LHC can easily provide the center-of-mass energies considered here. However, since linear colliders provide a cleaner environment with less background processes and underlying events, a linear collider is assumed in this study. 

Event generation is performed for each scenario independently and both beams are assumed to be unpolarised. The beamstrahlung effects \cite{beamstrahlung} are taken into account and the detector response is simulated based on the SiD detector at the International Linear Collider (ILC) \cite{SiDatILC}. Simulated signal and relevant background events are analyzed to reconstruct Higgs bosons by first performing jet clustering and $b$-tagging and then applying proper selection cuts to enrich the signal events. Identifying proper $b\bar{b}$ and $jjb\bar{b}$ combinations and computing their invariant masses, candidate mass distributions of the Higgs bosons are obtained at the assumed integrated luminosities. The integrated luminosity is set to 500 $fb^{-1}$ for all of the benchmark scenarios except for one scenario for which the integrated luminosity of 1000 $fb^{-1}$ is assumed (for search for the $A$ Higgs boson). It will be shown that, in all of the scenarios under consideration, both of the Higgs bosons $H$ and $A$ are observable with signals exceeding $5\sigma$. Moreover, reconstructing masses of the Higgs bosons, it will be shown that masses of both of the Higgs bosons are measurable. In what follows, a brief introduction to the 2HDM and its different types is provided, and then analysis and results will be discussed.

\section{Two-Higgs-doublet model} 
Employing two $SU(2)$ Higgs doublets, the potential 
\begin{equation}
  \begin{aligned}
     \mathcal{V} & =\, m_{11}^2\Phi_1^\dagger\Phi_1+m_{22}^2\Phi_2^\dagger\Phi_2
    -\Big[m_{12}^2\Phi_1^\dagger\Phi_2+\mathrm{h.c.}\Big]
    \\
    &+\frac{1}{2}\lambda_1\Big(\Phi_1^\dagger\Phi_1\Big)^2
    +\frac{1}{2}\lambda_2\Big(\Phi_2^\dagger\Phi_2\Big)^2
    +\lambda_3\Big(\Phi_1^\dagger\Phi_1\Big)\Big(\Phi_2^\dagger\Phi_2\Big)
    \\& +\lambda_4\Big(\Phi_1^\dagger\Phi_2\Big)\Big(\Phi_2^\dagger\Phi_1\Big)
    +\Big\{\frac{1}{2}\lambda_5\Big(\Phi_1^\dagger\Phi_2\Big)^2
    +\Big[\lambda_6\Big(\Phi_1^\dagger\Phi_1\Big)
     \\& +\lambda_7\Big(\Phi_2^\dagger\Phi_2\Big)
      \Big]\Big(\Phi_1^\dagger\Phi_2\Big)
    +\mathrm{h.c.}\Big\},
  \end{aligned}
  \label{lag}
\end{equation}
where $\Phi_1$ and $\Phi_2$ are $SU(2)$ Higgs doublets, is assumed as the Higgs potential in a general 2HDM. Using one additional Higgs doublet in this model leads to the prediction of five Higgs bosons. The lightest Higgs boson ($h$) is assumed to be the same as the observed Standard Model Higgs boson and the four additional Higgs bosons are thought of as yet undiscovered Higgs bosons. Additional Higgs bosons include a neutral scalar ($H$), a neutral pseudoscalar ($A$) and two charged ($H^\pm$) Higgs bosons. To respect experimental observations, the discrete $Z_2$ symmetry is imposed to avoid tree level flavor-changing neutral currents \cite{2hdm2,2hdm3,2hdm4_CompositeHiggs}. Consequently, it is implied that the parameters $\lambda_6$, $\lambda_7$ and $m_{12}^2$ must be zero. However, letting $m_{12}^2$ be non-zero, $Z_2$ symmetry is softly broken in this model. Assigning a value to $\tanb$ which is defined as the ratio of the vacuum expectation values of the two Higgs doublets, the parameters $m_{11}^2$ and $m_{22}^2$ are obtained by the minimization conditions for a minimum of the vacuum. Setting $\lambda_6$ and $\lambda_7$ to zero to respect the discrete $Z_2$ symmetry and working in the ``physical basis'', $\tanb$, mixing angle $\alpha$, $m_{12}^2$ and physical masses of the Higgs bosons must be determined to specify the model completely \cite{2hdm_TheoryPheno}. As a result of the imposed $Z_2$ symmetry, Higgs-fermion coupling scenarios are limited. Tab. \ref{coupling} provides Higgs coupling to up-type and down-type quarks and leptons in the allowed types which naturally conserve flavor. 
\begin{table}[h]
\normalsize
\fontsize{11}{7.2} 
    \begin{center}
         \begin{tabular}{ >{\centering\arraybackslash}m{.8in} >{\centering\arraybackslash}m{.4in} >{\centering\arraybackslash}m{.4in} >{\centering\arraybackslash}m{.4in}   }
& {$u_R^i$} & {$d_R^i$} & {$\ell_R^i$} \parbox{0pt}{\rule{0pt}{1ex+\baselineskip}}\\ \Xhline{3\arrayrulewidth}
 {Type \RN{1}} &$\Phi_2$ &$\Phi_2$ &$\Phi_2$ \parbox{0pt}{\rule{0pt}{1ex+\baselineskip}}\\ 
 {Type $\RN{2}$} &$\Phi_2$ &$\Phi_1$ &$\Phi_1$ \parbox{0pt}{\rule{0pt}{1ex+\baselineskip}}\\ 
{Type X} &$\Phi_2$ &$\Phi_2$ &$\Phi_1$  \parbox{0pt}{\rule{0pt}{1ex+\baselineskip}}\\ 
{Type Y} &$\Phi_2$ &$\Phi_1$ &$\Phi_2$  \parbox{0pt}{\rule{0pt}{1ex+\baselineskip}}\\ \Xhline{3\arrayrulewidth}
  \end{tabular}
\caption{Higgs coupling to up-type quarks, down-type quarks and leptons in types allowed by the $Z_2$ symmetry. The superscript $i$ is a generation index.  \label{coupling}}
  \end{center}
\end{table}
The types ``X'' and ``Y'' are also called ``lepton-specific'' and ``flipped'' respectively. Assuming $\sin(\bma)=1$ \cite{2hdm_TheoryPheno}, we choose the model to be SM-like in order for the $h$ Higgs boson to be thought of as the observed SM Higgs boson. Consequently, $h$ coupling to fermions reduce to corresponding couplings in the SM Yukawa Lagrangian. Consequently, the neutral Higgs part of the Yukawa Lagrangian takes the form \cite{2hdm_TheoryPheno,Barger_2hdmTypes}
\begin{equation}
\begin{aligned}
& \mathcal{L}_{ Yukawa}\, =\,  -v^{-1}  \Big(\, m_d\, \bar{d}d\, +\, m_u\, \bar{u}u\, +\, m_\ell\, \bar{\ell}\ell\, \Big)\ h \\
      & +v^{-1} \Big(\, \rho^dm_d\, \bar{d}d\, +\, \rho^um_u\, \bar{u}u\, +\, \rho^\ell m_\ell\, \bar{\ell}\ell\, \Big)\, H \\
& +iv^{-1}\Big(-\rho^dm_d\, \bar{d}\gamma_5d\, +\, \rho^um_u\, \bar{u}\gamma_5u\, -\, \rho^\ell m_\ell\, \bar{\ell}\gamma_5\ell\, \Big) A,
\label{yukawa2}
\end{aligned}
\end{equation}
 where $\rho^X$ factors are provided in Tab. \ref{rho}.
\begin{table}[h]
\normalsize
\fontsize{11}{7.2} 
    \begin{center}
         \begin{tabular}{ >{\centering\arraybackslash}m{.5in}  >{\centering\arraybackslash}m{.55in}  >{\centering\arraybackslash}m{.55in} >{\centering\arraybackslash}m{.55in}  >{\centering\arraybackslash}m{.55in}}
& {\RN{1}} & {$\RN{2}$} & {X} & {Y} \parbox{0pt}{\rule{0pt}{1ex+\baselineskip}}\\ \Xhline{3\arrayrulewidth}
 {$\rho^d$} &$\cot{\beta}$ &$- \tan\beta$ &$\cot\beta$ &$-\tan\beta$ \parbox{0pt}{\rule{0pt}{1ex+\baselineskip}}\\ 
{$\rho^u$} &$\cot{\beta}$ &$\cot\beta$ &$\cot\beta$ &$\cot\beta$  \parbox{0pt}{\rule{0pt}{1ex+\baselineskip}}\\   
{$\rho^\ell$} &$\cot{\beta}$ &$- \tan\beta$ &$-\tan\beta$&$\cot\beta$  \parbox{0pt}{\rule{0pt}{1ex+\baselineskip}}\\ \Xhline{3\arrayrulewidth} 
 \end{tabular}
\caption{$\rho^X$ factors corresponding to different flavor-conserving 2HDM types. \label{rho}} 
  \end{center}          
\end{table}   

As seen in Tab. \ref{rho}, coupling factors of different types differ dramatically leading to considerable differences in phenomenological features of different types \cite{2hdm_HiggsSector2}. According to Tab. \ref{rho}, factors of the Type-\RN{1} acquire the same values ($\cot\beta$) resulting in an interesting environment in both low and high $\tanb$ regions. Searching for the Higgs bosons in the Type-\RN{1} in this study is highly motivated by the large enhancement the $H\rightarrow b\bar{b}$ channel receives at low $\tanb$ values.

\section{Signal process}
The process chain $e^- e^+ \rightarrow A H\rightarrow ZHH \rightarrow jj b\bar{b}b\bar{b}$ is assumed as the signal process in this study and the Type-\RN{1} 2HDM is assumed as the theoretical framework to benefit from possible enhancements in low $\tanb$ regime. $jj$ is a pair of jets resulting from the $Z$ boson hadronic decay $Z\rightarrow jj$. After the pseudoscalar Higgs boson undergoes the decay channel $A\rightarrow ZH$, both of the scalar CP-even Higgs bosons experience the decay mode $H\rightarrow b\bar{b}$ which receives a large enhancement due to the $\cot\beta$ factor in the Higgs-fermion coupling factors as shown in Tab. \ref{rho} and thus, is dominant in low $\tanb$ regime as long as the scalar Higgs mass $m_H$ is below the threshold of the on-shell top quark pair production. The initial $e^-e^+$ collision is assumed to occur at the center-of-mass energies of $500$ and $1000$ GeV at a linear collider. 

Several benchmark points with different mass hypotheses are assumed in the parameter space of the 2HDM as shown in Tab. \ref{BPs}. 
\begin {table*}[!htbp]  
\begin{subtable}[b]{.29\textwidth}
\centering
\begin{tabular}{ccc} 
\multicolumn{3}{ c }{$\sqrt s=500$ GeV } \\ \Xhline{3\arrayrulewidth}
& BP1 & BP2 \\ \Xhline{3\arrayrulewidth}
$m_{h}$ & \multicolumn{2}{ c }{125} \\ 
$m_{H}$ & 150 & 150 \\ 
$m_{A}$ & 200 & 250  \\ 
$m_{H^\pm}$ & 200 & 250  \\ 
$m_{12}^2$ & 1987-2243 & 1987-2243 \\   
$\tan\beta$ & \multicolumn{2}{ c }{10} \\ 
$\sin(\beta-\alpha)$ & \multicolumn{2}{ c }{1} \\ 
$\sigma$ $[fb]$ & 3.8 & 2.9 \\ \Xhline{3\arrayrulewidth}
\end{tabular}
\caption {}
\label{BPs500}
\end{subtable}   
\begin{subtable}[b]{.64\textwidth}
\centering
\begin{tabular}{ccccccc} 
\multicolumn{7}{ c }{$\sqrt s=1000$ GeV } \\ \Xhline{3\arrayrulewidth}
& BP1 & BP2 & BP3 & BP4 & BP5 & BP6\\ \Xhline{3\arrayrulewidth}
$m_{h}$ & \multicolumn{6}{ c }{125} \\ 
$m_{H}$ & 150 & 150 & 200 & 200 & 250 & 250 \\
$m_{A}$ & 200 & 250 & 250 & 300 & 300 & 330 \\ 
$m_{H^\pm}$ & 200 & 250 & 250 & 300 & 300 & 330 \\ 
$m_{12}^2$ & 1987-2243 & 1987-2243  & 3720-3975 & 3720-3975 & 5948-6203 & 5948-6203 \\  
$\tan\beta$ & \multicolumn{6}{ c }{10} \\ 
$\sin(\beta-\alpha)$ & \multicolumn{6}{ c }{1} \\ 
$\sigma$ $[fb]$ & 2.8 & 3.7 & 1.4 & 2.1 & 0.5 & 0.8 \\ \Xhline{3\arrayrulewidth}
\end{tabular}
\caption {}
\label{BPs1000} 
\end{subtable}  
\caption {Benchmark points assumed for the center-of-mass energies of a) $500$ and b) $1000$ GeV. $\mh,\mH, \ma$ and $\mhpm$ are physical masses of the Higgs bosons and the provided $m^2_{12}$ range is the range satisfying theoretical requirements. The signal cross section is also provided for each scenario. \label{BPs}}
\end {table*} 
Benchmark points corresponding to the two assumed center-of-mass energies are simulated and analyzed separately. The scalar Higgs boson mass $m_H$ is assumed to vary in range $150$-$250$ GeV and the mass splitting between the $H$ and $A$ Higgs bosons is assumed to range from $50$ to $100$ GeV. $\tanb$ is set to 10 in all scenarios resulting in a considerable enhancement in the assumed scalar Higgs boson decay channel. 

To ensure that the considered scenarios are consistent with theoretical constraints, potential stability \cite{Deshpande}, perturbativity and unitarity \cite{Huffel,Maalampi,KANEMURA,GAKEROYD} of each scenario is checked with the use of \texttt{2HDMC 1.7.0} \cite{2hdmc1,2hdmc2} and the allowed range for $m^2_{12}$ parameter is provided in Tab. \ref{BPs}. Moreover, the assumed masses of the Higgs bosons are consistent with results of 86 analyses as checked by \texttt{HiggsBounds 4.3.1} \cite{HiggsBounds4.3.1} and \texttt{HiggsSignals 1.3.0} \cite{HiggsSignals1.3.0}. 

The experimental constraint \cite{BERTOLINI,DENNER}, based on the measurement performed at LEP \cite{Yao}, limits the deviation of the parameter $\rho=m_W^2(m_Z\cos\theta_W)^{-2}$ from its SM value. To respect this constraint, the Higgs bosons $A$ and $H^\pm$ are assumed to have the same masses in all of the benchmark points. This is because of the demonstration provided in \cite{drho,Gerard:2007kn} that concludes that if any of the conditions
\begin{equation} 
m_A=m_{H^\pm},\,\,\, m_H=m_{H^\pm},
\label{negligibledrho}  
\end{equation}
is met, the deviation of the $\rho$ parameter from its SM value is negligible. As reported in \cite{ATLAS-2HDM-2}, the LHC experiment has excluded mass regions $m_A=310-410,\,335-400,\,350-400$ GeV for $m_H=150,\,200, \,250$ GeV respectively at $\tb=10$ in the Type-\RN{1}. The LHC experiments \cite{CMS-2HDM-2,ATLAS-2HDM-1} also put the limit $m_A>350$ on the pseudoscalar Higgs mass for $\tb<5$ in the Type-\RN{1}. Moreover, another LHC experiment \cite{ATLAS-2HDM-3} excludes the mass range $m_H=170-360$ GeV for $\tb<1.5$ in this type. It is obvious from Tab. \ref{BPs} that the assumed scenarios are consistent with the current constraints resulting from direct LHC observations. 

In the context of the Type-\RN{2}, flavor physics data \cite{Misiak,Misiak:2017bgg} puts the constraints $m_{H^\pm}>570$ GeV ($\tb>2$) and $m_{H^\pm}>700$ GeV ($\tb<2$) on the charged Higgs mass. However, as indicated in \cite{Misiak:2017bgg}, no condition limits the Type-\RN{1} for $\tb>2$. Moreover, it is shown by a review of LHC, LEP and Tevatron results \cite{Moretti:2016qcc} that there is no exclusion around $\sin(\bma)=1$ for $m_{H/A/H^\pm} = 500$ GeV in the Type-\RN{1} 2HDM.

As indicated in \cite{lep1,lep2,lepexclusion2}, the constraints $m_A\geq93.4$ GeV and $m_{H^\pm}\geq78.6$ GeV must be met by the pseudoscalar and charged Higgs bosons in the MSSM. Also, the LHC experiments \cite{CMSNeutralHiggs,ATLASNeutralHiggs} has excluded the mass range $m_{A/H}=200-400$ GeV for $\tb\geq5$ in this model. However, the constraints imposed on the MSSM has no effect on the Type-\RN{1} 2HDM since their structures, Higgs-fermion coupling constants, parameter spaces, etc., are basically different. Finally, it is concluded that the assumed benchmark scenarios in this study are completely safe and consistent with the current experimental and theoretical constraints.

The assumed decay channel $A \rightarrow ZH$ in the signal process receives an enhancement due to the assumption $\sin(\bma)=1$ which is needed for the SM-like scenario, since the $A$-$Z$-$H$ vertex depends on $\sin(\bma)$ in the Type-\RN{1} 2HDM. Moreover, the assumed non-zero mass splitting between $A$ and $H$ Higgs bosons in the range $50$-$100$ GeV facilitates the possibilities for this decay channel. On average, we obtain BR($A \rightarrow ZH$) $\simeq 0.72$ for the assumed benchmark points of Tab. \ref{BPs} using \texttt{2HDMC 1.7.0}. The $A$-$Z$-$H$ vertex appears twice in the signal process. First in the production process $e^+e^- \rightarrow Z^* \rightarrow HA$ and then in the assumed decay mode for the pseudoscalar Higgs boson. Therefore, due to the assumption $\sin(\bma)=1$, the signal process is independent of $\tanb$ as long as the scalar Higgs decay mode $H\rightarrow b\bar{b}$ is not considered. This fact provides an opportunity for the signal to benefit from the enhancement received by the decay channel $H\rightarrow b\bar{b}$ at low $\tanb$ values, since lowering $\tanb$ value doesn't result in any decrease in the cross section of the process $e^+e^- \rightarrow Z^* \rightarrow AH\rightarrow ZHH$. Setting the value of 10 for $\tanb$ for all of the scenarios, on average, we obtain BR($H \rightarrow b\bar{b}$) $\simeq 0.61$ for the assumed benchmark points using \texttt{2HDMC 1.7.0}. The $b$ quark pairs resulting from the $H$ decays annihilate into hadronic jets and are identified by first performing jet reconstruction and then performing a proper $b$-tagging algorithm. The identified $b$-jets are then used to reconstruct the $H$ Higgs bosons. Although choosing the hadronic decay channel $Z\rightarrow jj$ for the $Z$ boson may give rise to more errors in the final results due to the uncertainties arising from jet reconstruction and $b$-tagging algorithms, the hadronic decay channel is chosen since the large branching ratio of this channel (BR$_{\, Z \rightarrow jj}$ $\simeq 0.69$) may fully compensate for the potential arising errors. Reconstructing the $Z$ boson by the identified jets, the $A$ Higgs boson is also reconstructed with the help of the reconstructed $H$ Higgs boson.
 
Background processes relevant to the considered signal process include $W^\pm$ pair production, $Z/\gamma$ production, $Z$ pair production and top quark pair production. Cross sections of the signal and relevant background processes are computed at the center-of-mass energies of 500 and 1000 GeV by \texttt{CompHEP 4.5.2} \cite{CompHEP4.4} and are provided in tables \ref{BPs} and \ref{bgXsec}. According to table \ref{BPs}, observing scenarios with heavier Higgs masses must be more challenging since the Higgs masses and cross section behave oppositely to each other.  
\begin {table}[!htbp]
\begin{center}    
\begin{tabular}{ccccc} 
& $t\bar{t}$ & $W^+W^-$ & $ZZ$ & $Z/\gamma$ \\ \Xhline{3\arrayrulewidth}
$ \sigma$ $[fb]$ ($\sqrt s=500$ GeV) & 562 & 7887 & 450 & 16846 \\
$\sigma$ $[fb]$ ($\sqrt s=1000$ GeV) & 226 & 3410 & 190 & 4335  \\ \Xhline{3\arrayrulewidth}
\end{tabular}
\caption {Background cross sections. }
\label{bgXsec}
\end{center} 
\end {table}     
         
\section{Event generation}   
In order to take the beamstrahlung effects into account and simulate the detector response, event generation is performed in several steps for each benchmark scenario. Hard scattering part of the signal and background events (parton-level part which doesn't include parton showers, hadronization, etc.) are generated using \texttt{CompHEP 4.5.2}. Simulation of the beamstrahlung effects is also performed by \texttt{CompHEP} with the use of the beam parameters provided in Tab \ref{beamstrahlungParameters} and the assumption that both beams are unpolarised. 
\begin {table}[!htbp] 
\begin{center}   
\begin{tabular}{ccc}  
& $500$ GeV & $1000$ GeV \\ \Xhline{3\arrayrulewidth}
RMS horizontal beam size (nm) & 474 & 335 \\
RMS vertical beam size (nm) & 5.9 & 2.7 \\ 
RMS bunch length (mm) & 0.3 & 0.225 \\
No. of particles / bunch ($\times 10^{10}$) & 2 & 1.74 \\ \Xhline{3\arrayrulewidth}
\end{tabular}
\caption {Beam parameters corresponding to the center-of-mass energies of 500 and 1000 GeV taken from Tab. 8.2 of ILC technical design report v3.II \cite{ILCtdrv3.II} }
\label{beamstrahlungParameters}
\end{center}
\end {table}   
To generate the remaining part of the events including multi-particle interactions, parton showers and hadronization, the SLHA (SUSY Les Houches Accord) files generated by \texttt{2HDMC 1.7.0} as well as the parton-level events generated by \texttt{CompHEP} are passed to \texttt{PYTHIA 8.2.15} \cite{pythia82}. The SLHA files contain basic parameters of the Type-\RN{1} 2HDM including coupling constants, branching ratios, etc. Using the input files, \texttt{PYTHIA} performs further processing to complete the events. The generated events are then internally used by \texttt{DELPHES 3.4} \cite{DELPHES3.4} to simulate detector response with the use of the DSiD detector card which is based on the full simulation performance of the SiD detector at the ILC. The anti-$k_t$ algorithm \cite{antikt} from FASTJET 3.1.0 package \cite{fastjet1,fastjet2} is used to perform jet reconstruction with the cone size $\Delta R=\sqrt{(\Delta\eta)^2+(\Delta\phi)^2}=0.4$, where $\eta=-\textnormal{ln}\tan(\theta/2)$ and $\phi$ ($\theta$) is the azimuthal (polar) angle with respect to the beam axis. \texttt{DELPHES} output data is stored as \texttt{ROOT} files \cite{root1} and contains reconstructed jets and also $b$-tagging flags by which $b$-jets are identified. 

\section{Event selection and analysis}   
\begin{figure}[!htbp]
  \centering
    \begin{subfigure}[b]{0.48\textwidth}
    \centering
    \includegraphics[width=\textwidth]{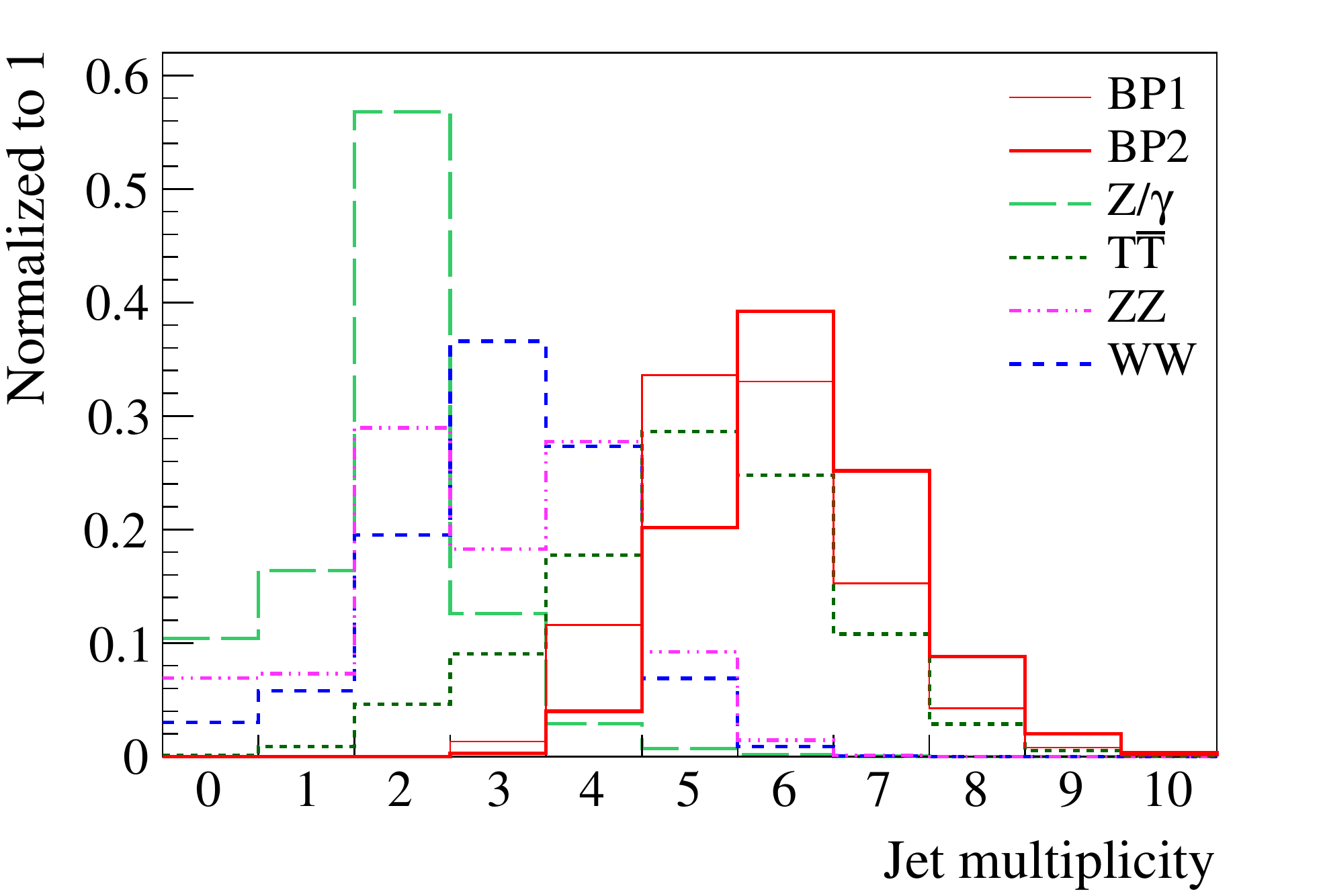}
    \caption{}
    \label{hnjets-500} 
    \end{subfigure}  
        \quad     
    \begin{subfigure}[b]{0.48\textwidth}
    \centering
    \includegraphics[width=\textwidth]{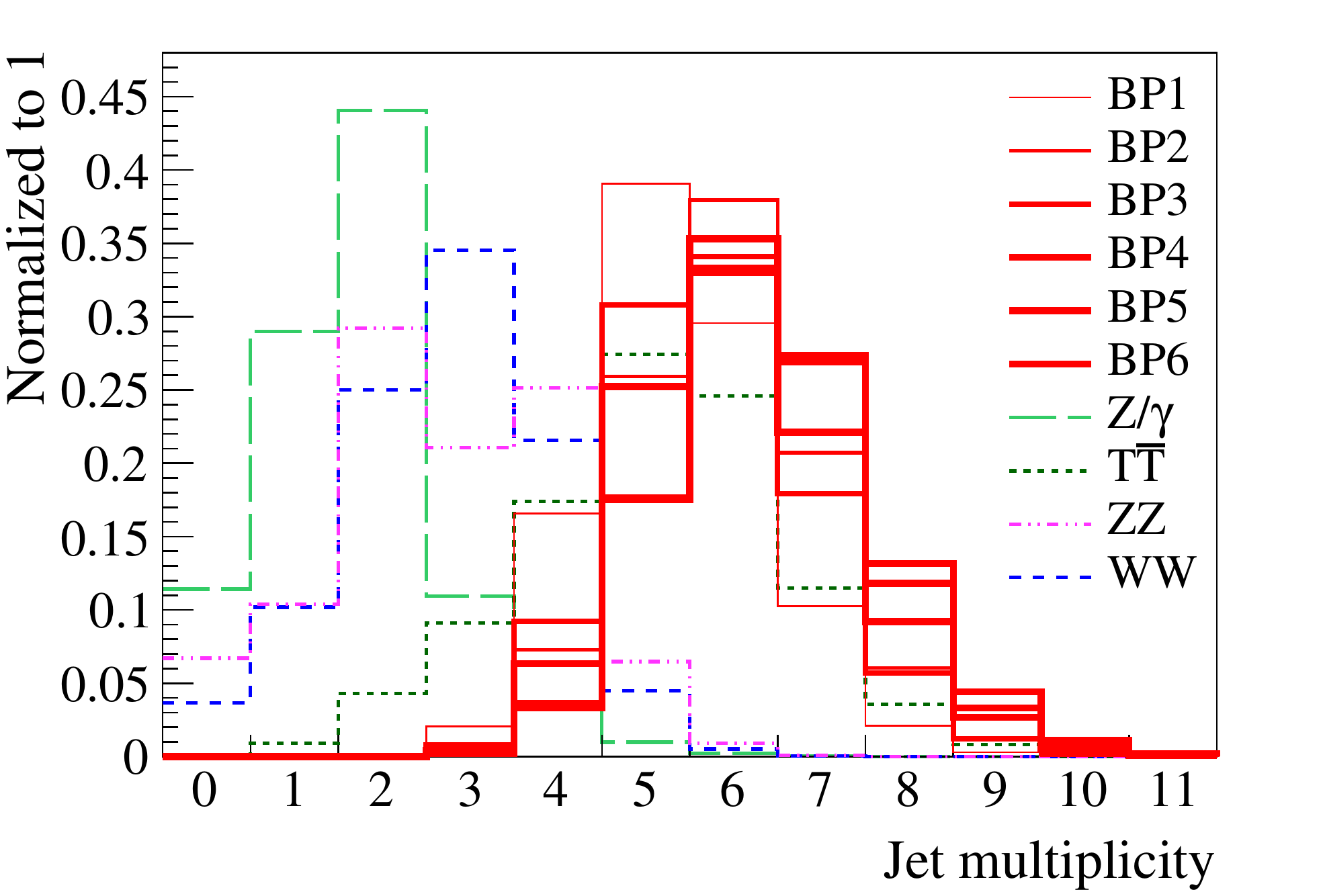}
    \caption{}
    \label{hnjets-1000}
    \end{subfigure}
  \caption{Jet multiplicity distributions corresponding to different signal and background processes at the center-of-mass energies of a) 500 and b) 1000 GeV.}
  \label{hnjets} 
\end{figure}

\begin{figure}[!htbp]
  \centering
    \begin{subfigure}[b]{0.48\textwidth}
    \centering 
    \includegraphics[width=\textwidth]{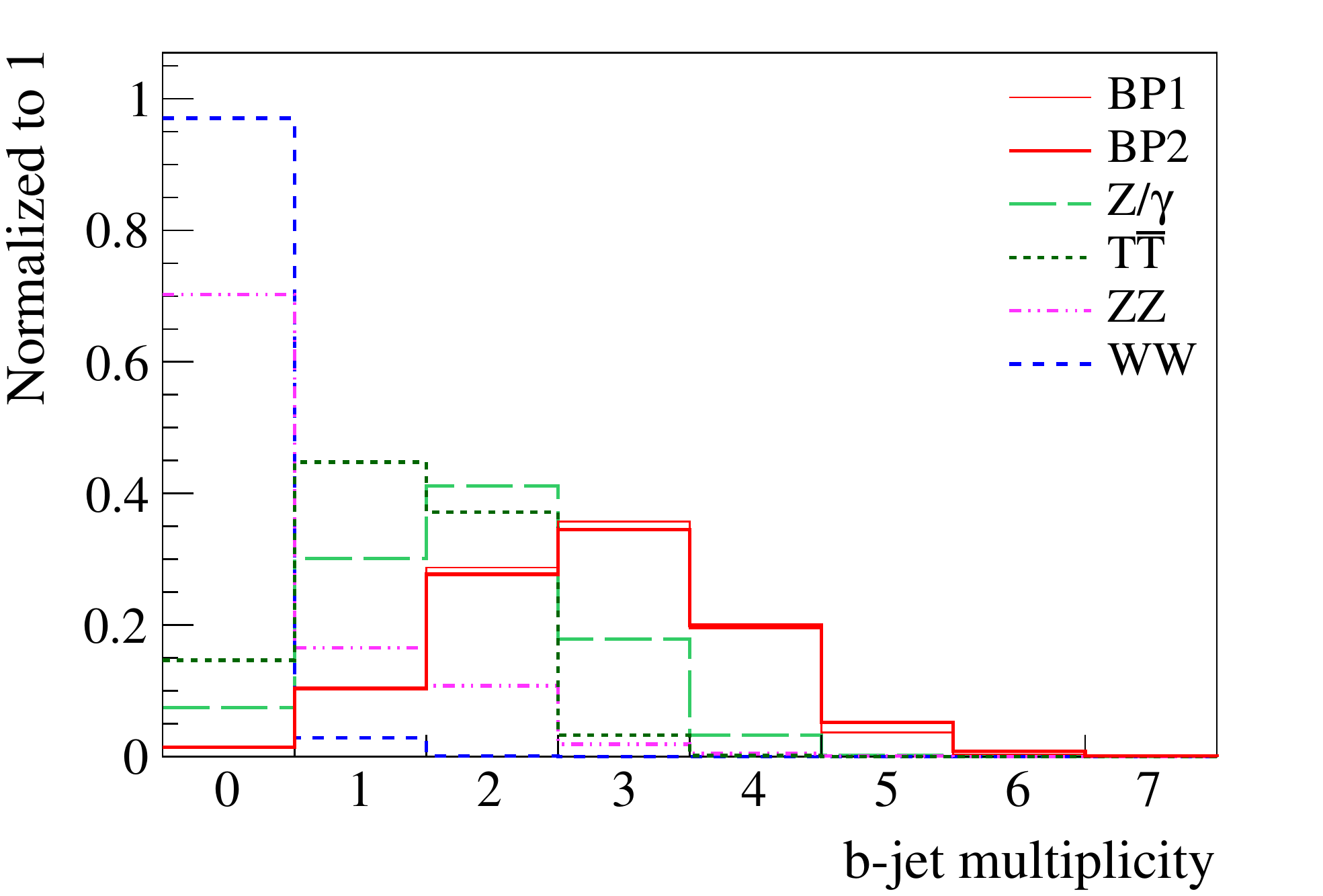}
    \caption{}
    \label{hnbjets-500} 
    \end{subfigure}  
        \quad    
    \begin{subfigure}[b]{0.48\textwidth}
    \centering
    \includegraphics[width=\textwidth]{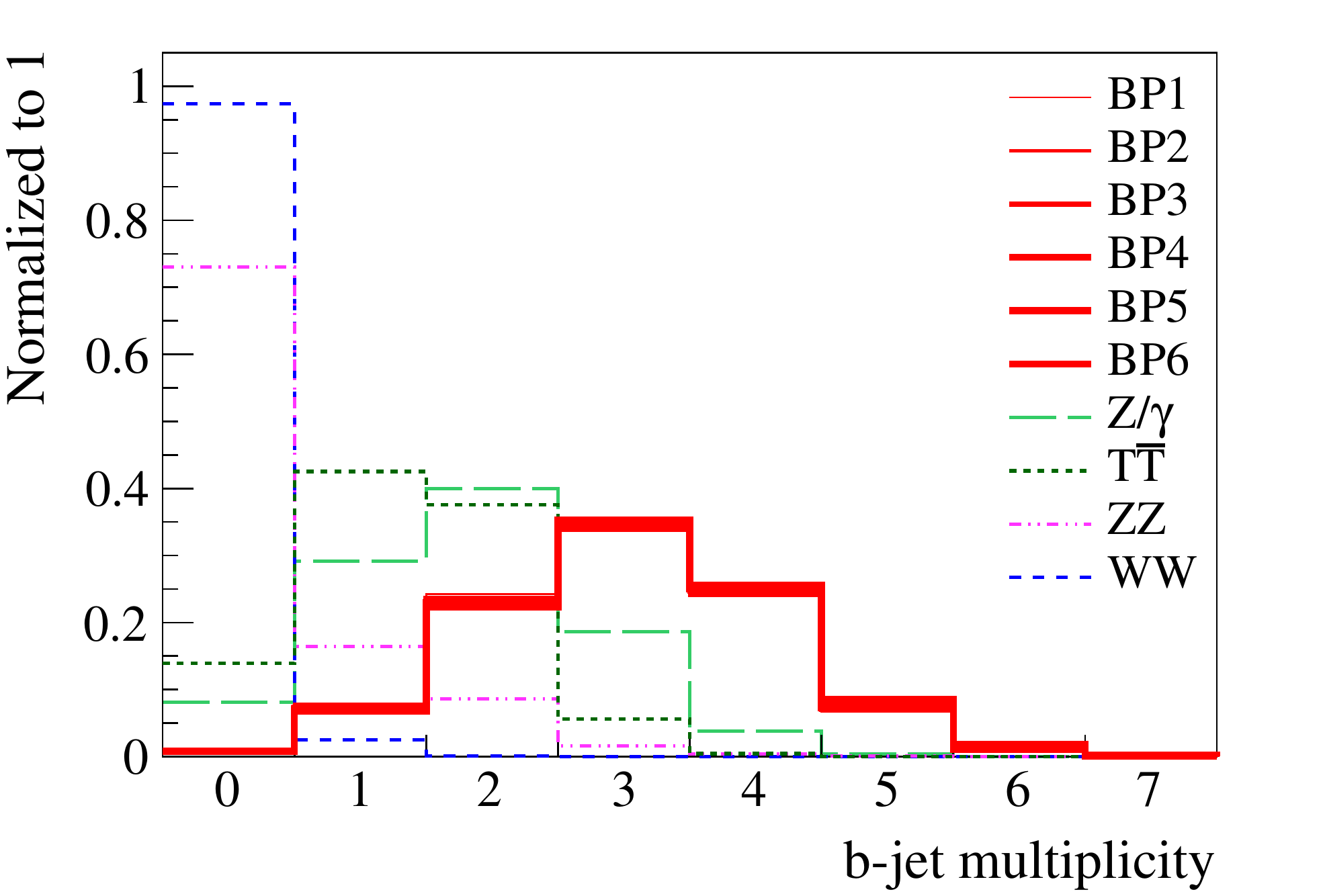}
    \caption{}
    \label{hnbjets-1000}
    \end{subfigure}
  \caption{$b$-jet multiplicity distributions corresponding to different signal and background processes at the center-of-mass energies of a) 500 and b) 1000 GeV.}
  \label{hnbjets}  
\end{figure}

Analysis begins by counting the number of reconstructed jets satisfying the kinematic thresholds
\begin{equation}
\begin{aligned}
& \bm{{p_T}}_{\bm{jet}}\geq10\ GeV,\ \ \  \vert \bm{\eta}_{\bm{jet}} \vert \leq 3, \\ 
\end{aligned}
\label{jetsconditions} 
\end{equation}  
where $p_T$ is the transverse momentum. Based on the jet multiplicity distributions of Fig. \ref{hnjets} 
which are obtained for the two assumed center-of-mass energies, the selection cut
\begin{equation}
\bm{N}_{\textbf{\emph{jet}}} \geq 5,
\label{jetsnumbercondition} 
\end{equation}
where $N_{\textbf{\emph{jet}}}$ is the number of jets, is applied. Using $b$-tagging flags, $b$-jets are identified and the $b$-jet multiplicity distributions of Fig. \ref{hnbjets} are obtained for the $b$-jets satisfying the lower threshold $\bm{{p_T}}_{\textbf{\emph{b-jet}}}\geq20\ GeV$.
The condition
\begin{equation}
\bm{N}_{\textbf{\emph{b-jet}}} \geq 3,
\label{bjetsnumbercondition} 
\end{equation}
where $N_{\textbf{\emph{b-jet}}}$ is the number of $b$-jets, is then applied to rule out events with less than three $b$-jets.
\begin{table*}[!htbp] 
\begin{subtable}[b]{.37\textwidth}
\normalsize
\fontsize{11}{7.2} 
    \begin{center}
        \begin{tabular}{ >{\centering\arraybackslash}m{.9in} >{\centering\arraybackslash}m{.5in}  >{\centering\arraybackslash}m{.5in}}
\multicolumn{3}{ c }{$\sqrt s=500$ GeV } \\ \Xhline{3\arrayrulewidth}
  & {BP1} & {BP2}  \parbox{0pt}{\rule{0pt}{1ex+\baselineskip}}\\ \Xhline{3\arrayrulewidth}
    {$N_{jet}\geq5$} & 0.870 & 0.957  \parbox{0pt}{\rule{0pt}{1ex+\baselineskip}}\\ 
   {${N}_{{\emph{b-jet}}} \geq 3$} & 0.593 & 0.605  \parbox{0pt}{\rule{0pt}{1ex+\baselineskip}}\\ 
   $N_{b\bar{b}}\geq 1$ & 0.576 & 0.590  \parbox{0pt}{\rule{0pt}{1ex+\baselineskip}}\\ 
    {\textbf{Total eff.}} & \textbf{0.297} & \textbf{0.342}  \parbox{0pt}{\rule{0pt}{1ex+\baselineskip}}\\ \Xhline{1\arrayrulewidth}
$N_{ZH}=1$ & 0.500 & 0.604  \parbox{0pt}{\rule{0pt}{1ex+\baselineskip}}\\ 
    {\textbf{Total eff.}} & \textbf{0.149} & \textbf{0.207}  \parbox{0pt}{\rule{0pt}{1ex+\baselineskip}}\\ \Xhline{3\arrayrulewidth}
        \end{tabular}
\caption{} 
\label{signaleff500}      
  \end{center}
\end{subtable} 
\begin{subtable}[b]{.54\textwidth}
\normalsize
\fontsize{11}{7.2} 
    \begin{center}
        \begin{tabular}{>{\centering\arraybackslash}m{.88in}  >{\centering\arraybackslash}m{.4in}  >{\centering\arraybackslash}m{.4in} >{\centering\arraybackslash}m{.4in} >{\centering\arraybackslash}m{.4in} >{\centering\arraybackslash}m{.4in} >{\centering\arraybackslash}m{.44in}}
\multicolumn{7}{ c }{$\sqrt s=1000$ GeV } \\ \Xhline{3\arrayrulewidth}
  & {BP1} & {BP2} & {BP3} & {BP4} & {BP5} & {BP6} \parbox{0pt}{\rule{0pt}{1ex+\baselineskip}}\\ \Xhline{3\arrayrulewidth}
    {$N_{jet}\geq5$} & 0.813 & 0.920 & 0.900 & 0.961 & 0.932 & 0.965 \parbox{0pt}{\rule{0pt}{1ex+\baselineskip}}\\ 
   {$N_{\emph{b-jet}}\geq3$} & 0.669 & 0.675 & 0.680 & 0.694 & 0.692 & 0.701 \parbox{0pt}{\rule{0pt}{1ex+\baselineskip}}\\ 
   $N_{b\bar{b}}\geq 1$ & 0.765 & 0.812 & 0.899 & 0.904 & 0.911 & 0.911 \parbox{0pt}{\rule{0pt}{1ex+\baselineskip}}\\ 
    {\textbf{Total eff.}} & \textbf{0.416} & \textbf{0.505} & \textbf{0.550} & \textbf{0.603} & \textbf{0.588} & \textbf{0.616} \parbox{0pt}{\rule{0pt}{1ex+\baselineskip}}\\  \Xhline{1\arrayrulewidth}
   {$N_{ZH}=1$} & 0.208 & 0.282 & 0.276 & 0.330 & 0.300 & 0.321 \parbox{0pt}{\rule{0pt}{1ex+\baselineskip}}\\ 
    {\textbf{Total eff.}} & \textbf{0.087} & \textbf{0.142} & \textbf{0.152} & \textbf{0.199} & \textbf{0.176} & \textbf{0.198} \parbox{0pt}{\rule{0pt}{1ex+\baselineskip}}\\ \Xhline{3\arrayrulewidth}
        \end{tabular}
\caption{}
\label{signaleff1000}
  \end{center}
\end{subtable}
\newline\newline
\begin{subtable}[b]{.49\textwidth} 
\normalsize
\fontsize{11}{7.2} 
    \begin{center}
        \begin{tabular}{ >{\centering\arraybackslash}m{.9in} >{\centering\arraybackslash}m{.5in}  >{\centering\arraybackslash}m{.5in} >{\centering\arraybackslash}m{.5in}  >{\centering\arraybackslash}m{.5in}}
\multicolumn{5}{ c }{$\sqrt s=500$ GeV } \\ \Xhline{3\arrayrulewidth}
  & {$t\bar{t}$} & {$WW$} & {$ZZ$} & {$Z/\gamma$ } \parbox{0pt}{\rule{0pt}{1ex+\baselineskip}}\\ \Xhline{3\arrayrulewidth}
    {$N_{jet}\geq5$} & 0.677 & 0.078 & 0.108 & 0.009 \parbox{0pt}{\rule{0pt}{1ex+\baselineskip}}\\ 
   {${N}_{{\emph{b-jet}}} \geq 3$} & 0.034 & 7e-05 & 0.024 & 0.214  \parbox{0pt}{\rule{0pt}{1ex+\baselineskip}}\\  
   $N_{b\bar{b}}\geq 1$ & 0.117 & 0.018 & 0.264 & 0.129 \parbox{0pt}{\rule{0pt}{1ex+\baselineskip}}\\ 
    {\textbf{Total eff.}} & \textbf{0.003} & \textbf{1e-07} & \textbf{7e-4} & \textbf{3e-4}  \parbox{0pt}{\rule{0pt}{1ex+\baselineskip}}\\ \Xhline{1\arrayrulewidth}
   $N_{ZH}=1$ & 0.498 & 0.500 & 0.065 & 0.245 \parbox{0pt}{\rule{0pt}{1ex+\baselineskip}}\\ 
    {\textbf{Total eff.}} & \textbf{0.001} & \textbf{5e-08} & \textbf{4e-05} & \textbf{6e-05}  \parbox{0pt}{\rule{0pt}{1ex+\baselineskip}}\\ \Xhline{3\arrayrulewidth}
        \end{tabular}
\caption{}
 \label{bkgeff500}
  \end{center}
\end{subtable} 
\begin{subtable}[b]{.49\textwidth} 
\normalsize
\fontsize{11}{7.2} 
    \begin{center}
        \begin{tabular}{ >{\centering\arraybackslash}m{.88in}  >{\centering\arraybackslash}m{.51in}  >{\centering\arraybackslash}m{.51in} >{\centering\arraybackslash}m{.51in} >{\centering\arraybackslash}m{.55in} }
\multicolumn{5}{ c }{$\sqrt s=1000$ GeV } \\ \Xhline{3\arrayrulewidth}
  & {$t\bar{t}$} & {$WW$} & {$ZZ$} & {$Z/\gamma$ } \parbox{0pt}{\rule{0pt}{1ex+\baselineskip}}\\ \Xhline{3\arrayrulewidth}
    {$N_{jet}\geq5$} & 0.681 & 0.050 & 0.075 & 0.013  \parbox{0pt}{\rule{0pt}{1ex+\baselineskip}}\\ 
   {$N_{\emph{b-jet}}\geq3$} & 0.061 & 1e-4 & 0.020 & 0.228 \parbox{0pt}{\rule{0pt}{1ex+\baselineskip}}\\ 
   $N_{b\bar{b}}\geq 1$ & 0.223 & 0.075 & 0.462 & 0.316 \parbox{0pt}{\rule{0pt}{1ex+\baselineskip}}\\ 
    {\textbf{Total eff.}} & \textbf{0.009} & \textbf{4e-07} & \textbf{7e-4} & \textbf{9e-4} \parbox{0pt}{\rule{0pt}{1ex+\baselineskip}}\\  \Xhline{1\arrayrulewidth}
   {$N_{ZH}=1$} & 0.090 & 0.125 & 0.030 & 0.127 \parbox{0pt}{\rule{0pt}{1ex+\baselineskip}}\\ 
    {\textbf{Total eff.}} & \textbf{8e-4} & \textbf{5e-08} & \textbf{2e-05} & \textbf{1e-4 } \parbox{0pt}{\rule{0pt}{1ex+\baselineskip}}\\ \Xhline{3\arrayrulewidth}
        \end{tabular}
\caption{}
\label{bkgeff1000} 
  \end{center}
\end{subtable}
\caption{Event selection efficiencies corresponding to a,b) signal and c,d) background processes assuming different benchmark scenarios at the center-of-mass energies of 500 and 1000 GeV.}
\label{eff}
\end{table*}
$b$-jets present in each events are analyzed to find the true $b$-jet pairs which originate from the $H$ Higgs bosons decays. In events with three $b$-jets, the value of $\Delta R_{\, bb}$, where $\Delta R$ follows the definition
\begin{equation} 
\Delta R=\sqrt{(\Delta\eta)^2+(\Delta\phi)^2},
\label{deltaR}
\end{equation}
 is computed for the three possible $b$-jet pairs and the pair which has the minimum $\Delta R_{\, bb}$ value is considered as the true $b$-jet pair if the condition 
\begin{equation}
 \bm{\Delta R_{\, bb}}\geq 
\begin{cases}
      1.7, & \sqrt s = 500\, \text{GeV} \\
      0.8, & \sqrt s = 1000\, \text{GeV}
    \end{cases}
 \label{dRbbcondition} 
\end{equation}  
is satisfied. In events with at least four $b$-jets, all possible combinations of two $b$-jet pairs are considered and the difference between the invariant masses of the two $b$-jet pairs $\vert M_{\, b_1b_2}-M_{\, b_3b_4}\vert$ is computed for each combination, and the $b$-jet pairs of the combination with minimum invariant mass difference are identified as true pairs if the condition \ref{dRbbcondition} is satisfied. Identifying true pairs, the selection cut
\begin{equation}  
\bm{N_{b\bar{b}}} \geq1,  
\label{Nbbcut} 
\end{equation} 
where $N_{b\bar{b}}$ is the number of identified true $b$-jet pairs, is imposed and the $H$ Higgs boson is reconstructed by the identified $b$-jet pairs. Counting the remaining jets which have not participated in the reconstruction of the $H$ Higgs boson, events with less than two remaining jets are vetoed to rule out events with no reconstructable $Z$ boson. Reconstructing the $Z$ boson by the remaining pair of jets, the $A$ Higgs boson can be reconstructed by the reconstructed $Z$ and $H$ bosons. At this stage, each event contains one or two reconstructed $H$ Higgs bosons and one reconstructed $Z$ boson. In events with one reconstructed $H$ Higgs boson, the combination $ZH$ is identified as the true combination if the condition
\begin{equation}
 \bm{\Delta R_{\, ZH}}\leq 
\begin{cases}
      2, & \sqrt s = 500\, \text{GeV} \\
      1, & \sqrt s = 1000\, \text{GeV}
    \end{cases}
 \label{dRZHcondition} 
\end{equation}  
where $\Delta R$ follows the definition of Eq. \ref{deltaR}, is satisfied. In events with two reconstructed $H$ Higgs bosons, $\Delta R_{ZH}$ is computed for each one of the two possible $ZH$ combinations and the combination with smaller $\Delta R_{ZH}$ value is identified as the true combination. Having true combinations identified, the selection cut
\begin{equation}  
\bm{N_{\, ZH}} =1,  
\label{Nbbcut} 
\end{equation} 
where $N_{\, ZH}$ is the number of identified true $ZH$ combinations, is applied and the $A$ Higgs boson is reconstructed using the identified $ZH$ combination. Applying the selection cuts, event selection efficiencies are obtained for different signal and background processes at $\sqrt s=$ 500 and 1000 GeV and the results are provided in Tab. \ref{eff}. According to the analysis, the Higgs bosons $H$ and $A$ are reconstructed after three and four selection cuts respectively. Thus, total efficiencies corresponding to the first three selection cuts and all of the four selection cuts are also provided. 

\begin{figure*}[h]
  \centering  
  \bigskip 
    \begin{subfigure}[b]{0.49\textwidth} 
    \centering 
     \includegraphics[width=\textwidth]{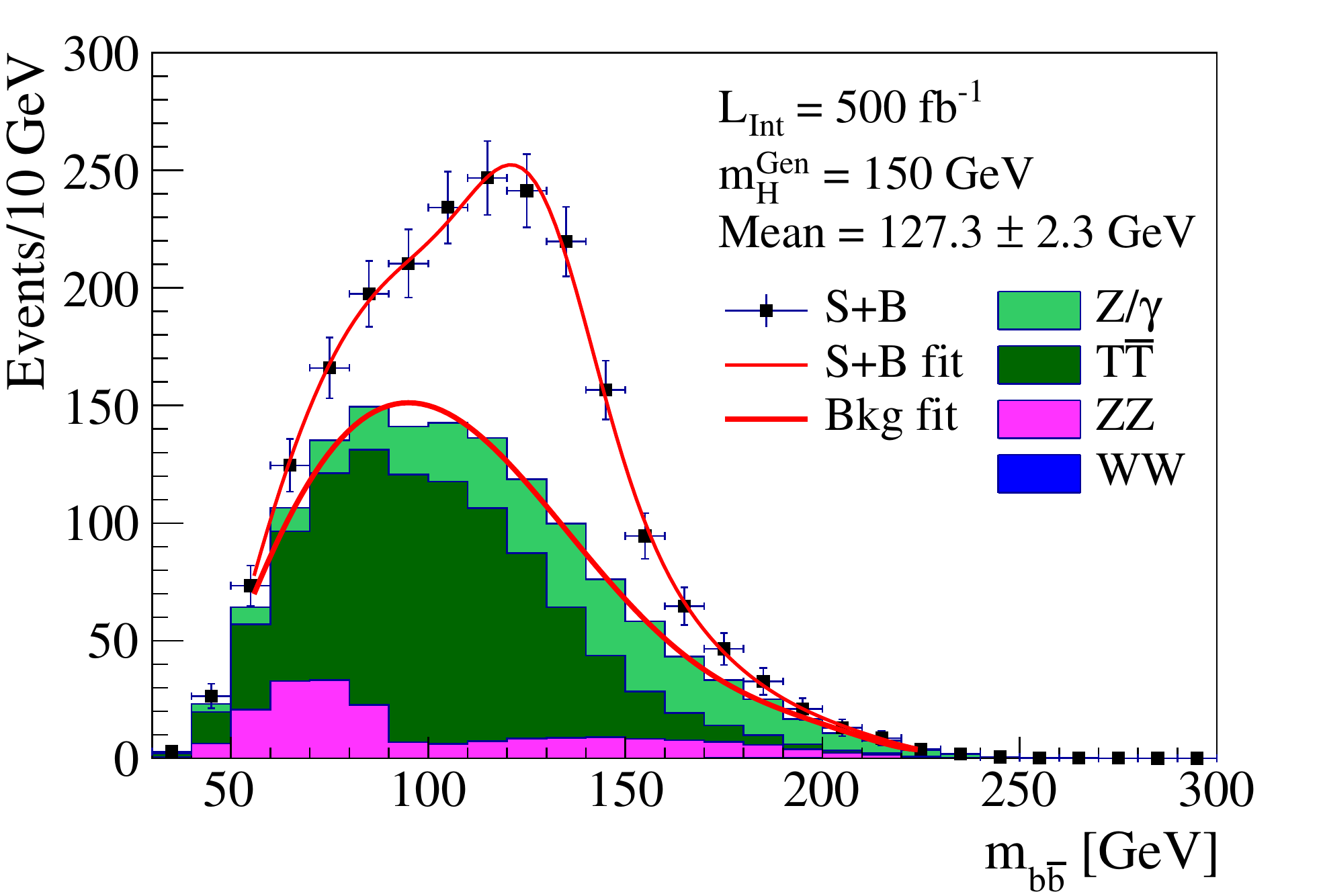}
    \caption{}
    \label{HBP1-500}  
    \end{subfigure}  
    \begin{subfigure}[b]{0.49\textwidth} 
    \centering 
    \includegraphics[width=\textwidth]{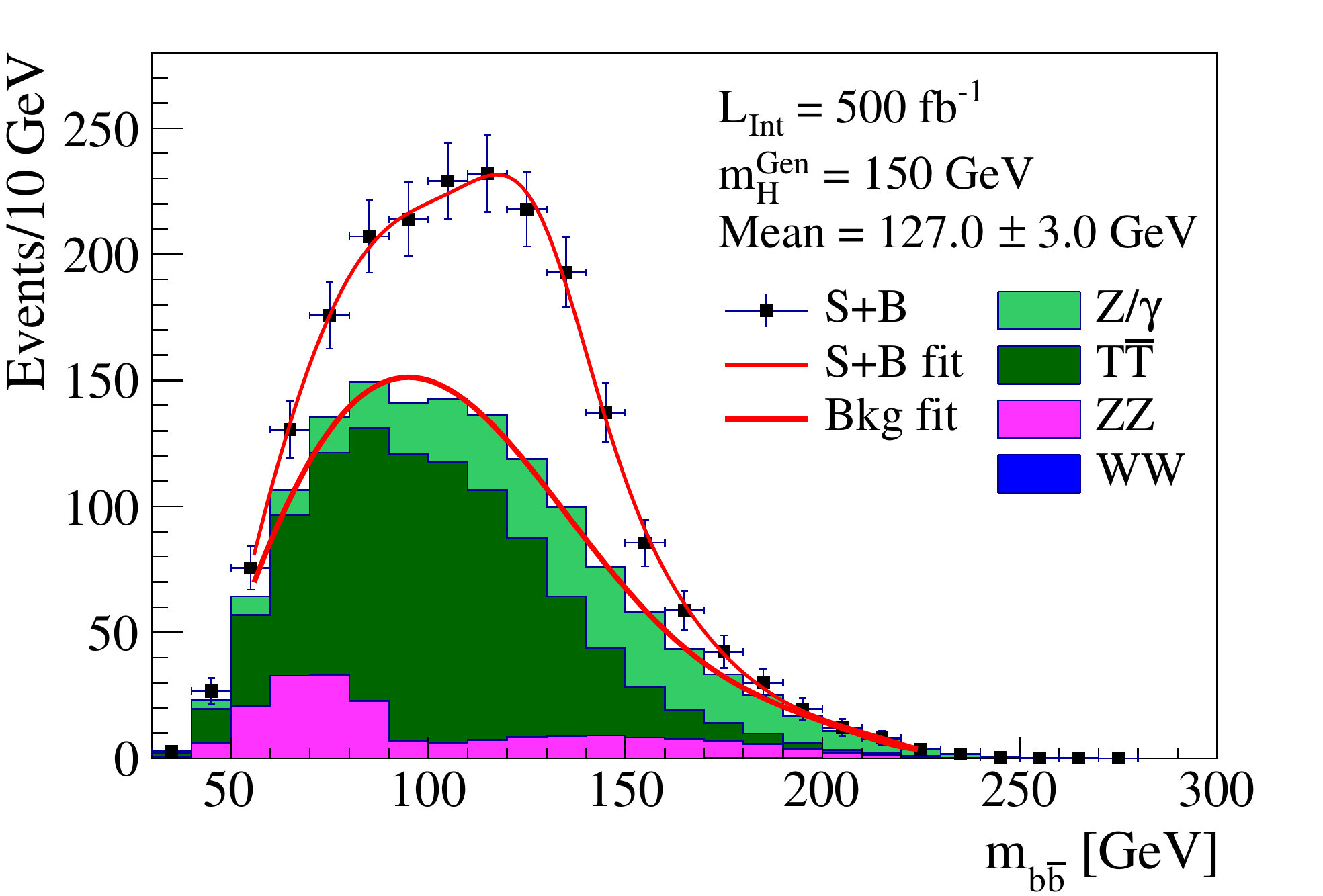}
    \caption{} 
    \label{HBP2-500} 
    \end{subfigure} 
    \begin{subfigure}[b]{0.49\textwidth} 
    \centering 
    \includegraphics[width=\textwidth]{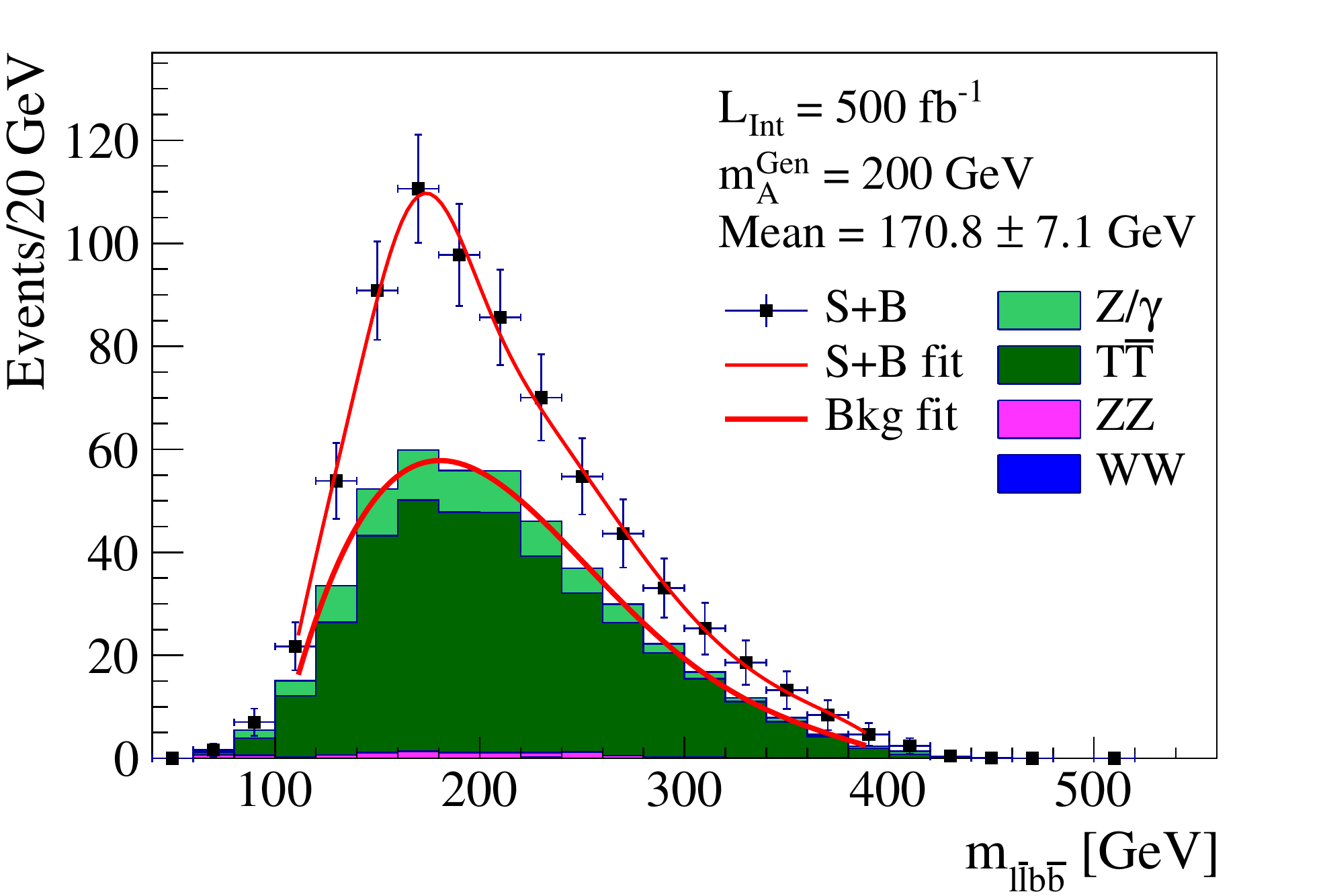}
    \caption{} 
    \label{ABP1-500} 
    \end{subfigure} 
    \begin{subfigure}[b]{0.49\textwidth} 
    \centering 
    \includegraphics[width=\textwidth]{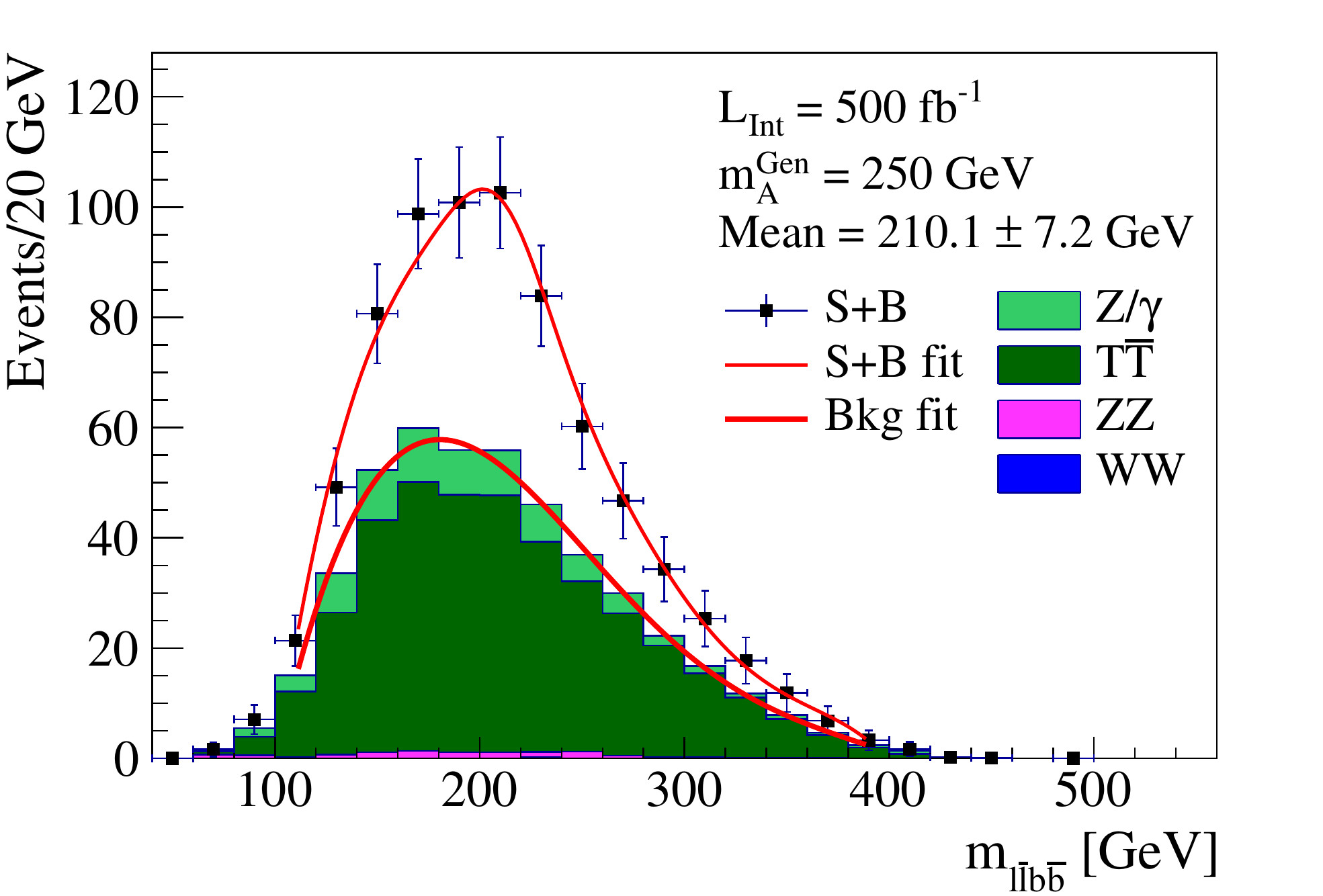}
    \caption{} 
    \label{ABP2-500} 
    \end{subfigure} 
\caption{Candidate mass distributions of the a,b) $H$ and c,d) $A$ Higgs bosons with corresponding fitting results and errors in different scenarios at $\sqrt s=500$ GeV.}
  \label{500fit} 
\end{figure*}   
\begin{figure*}[h]
  \centering  
    \begin{subfigure}[b]{0.49\textwidth} 
    \centering
    \includegraphics[width=\textwidth]{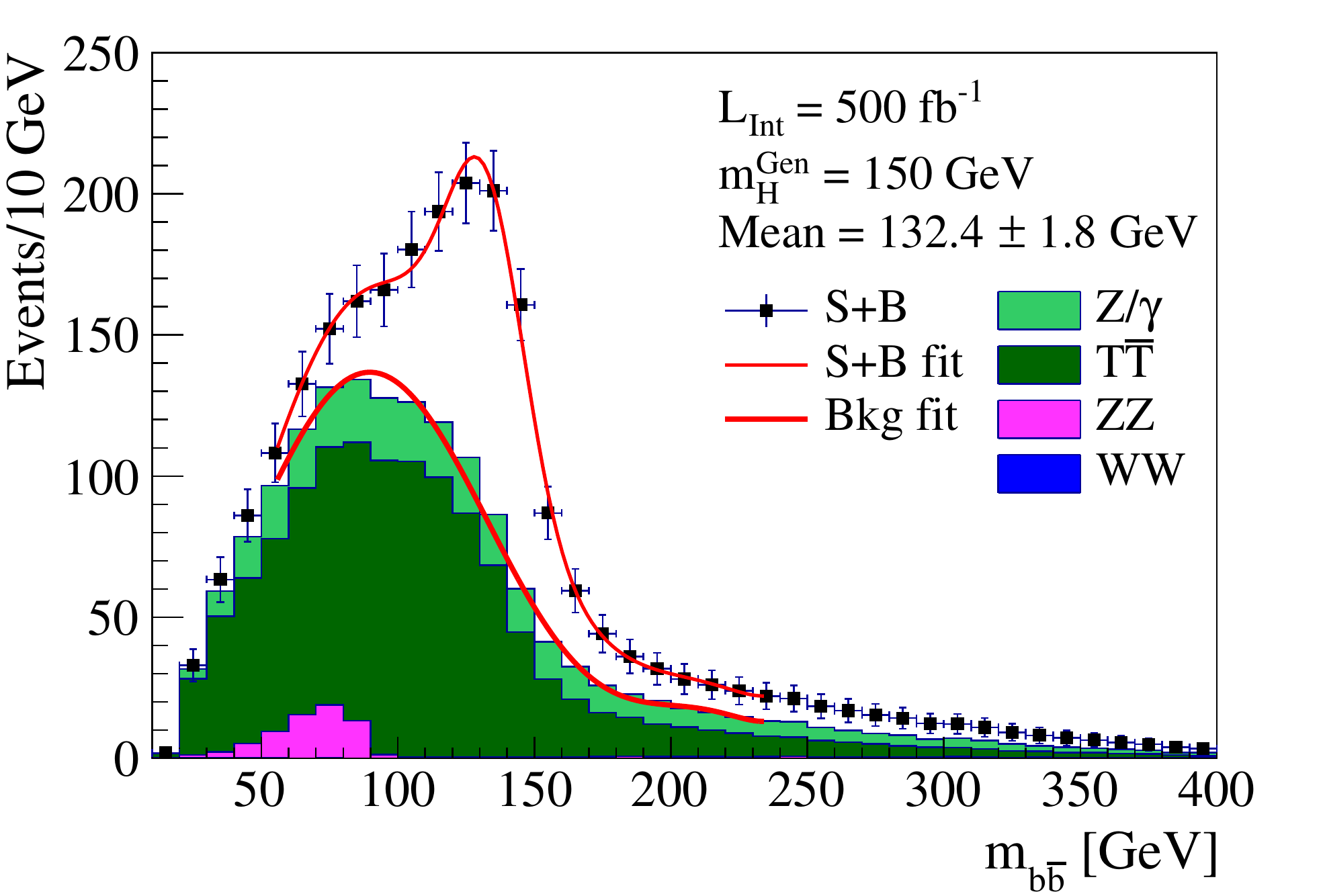}
    \caption{}
    \label{HBP1}
    \end{subfigure} 
    \begin{subfigure}[b]{0.49\textwidth}
    \centering
    \includegraphics[width=\textwidth]{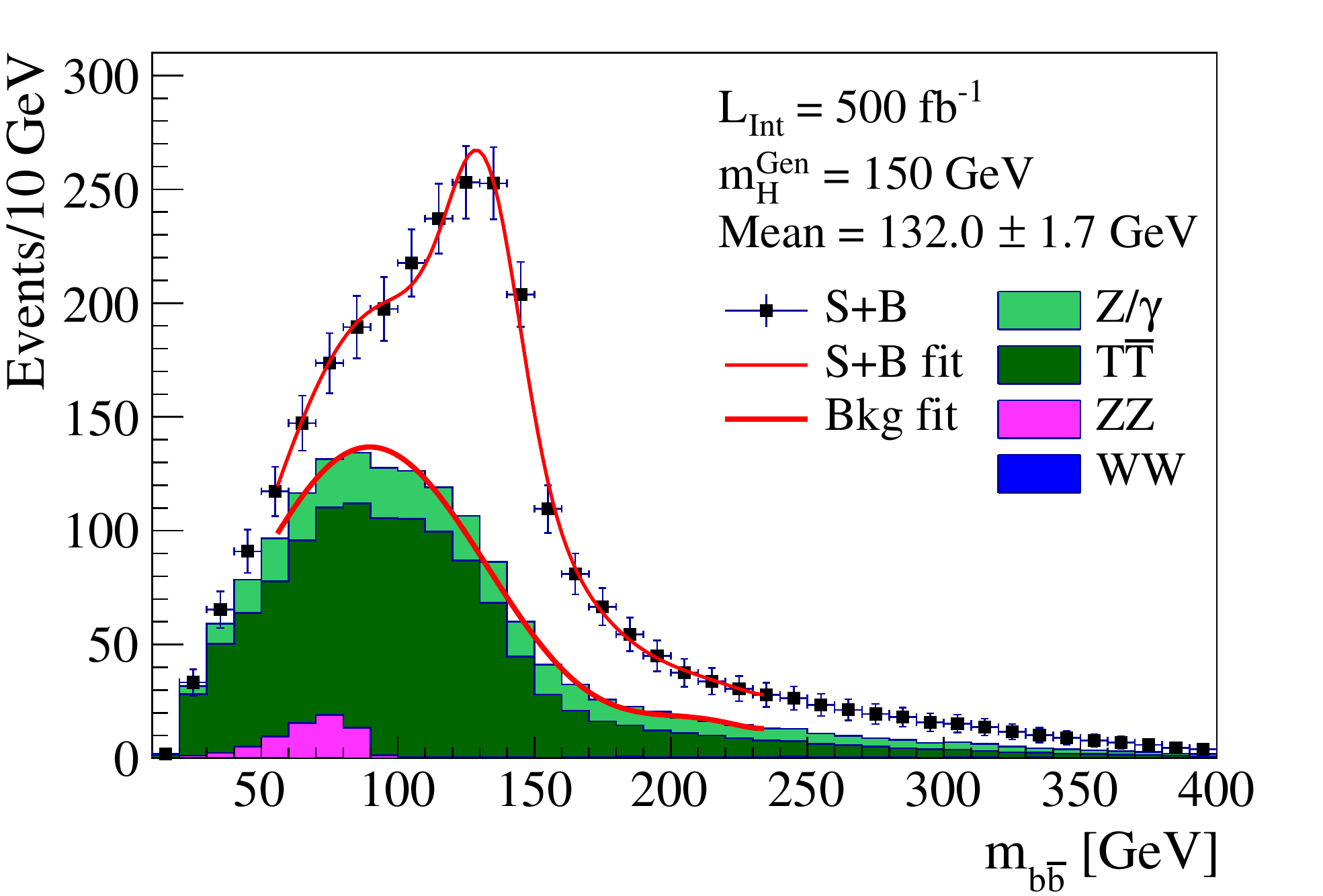}
    \caption{}
    \label{HBP2} 
    \end{subfigure} 
    \begin{subfigure}[b]{0.49\textwidth}
    \centering
    \includegraphics[width=\textwidth]{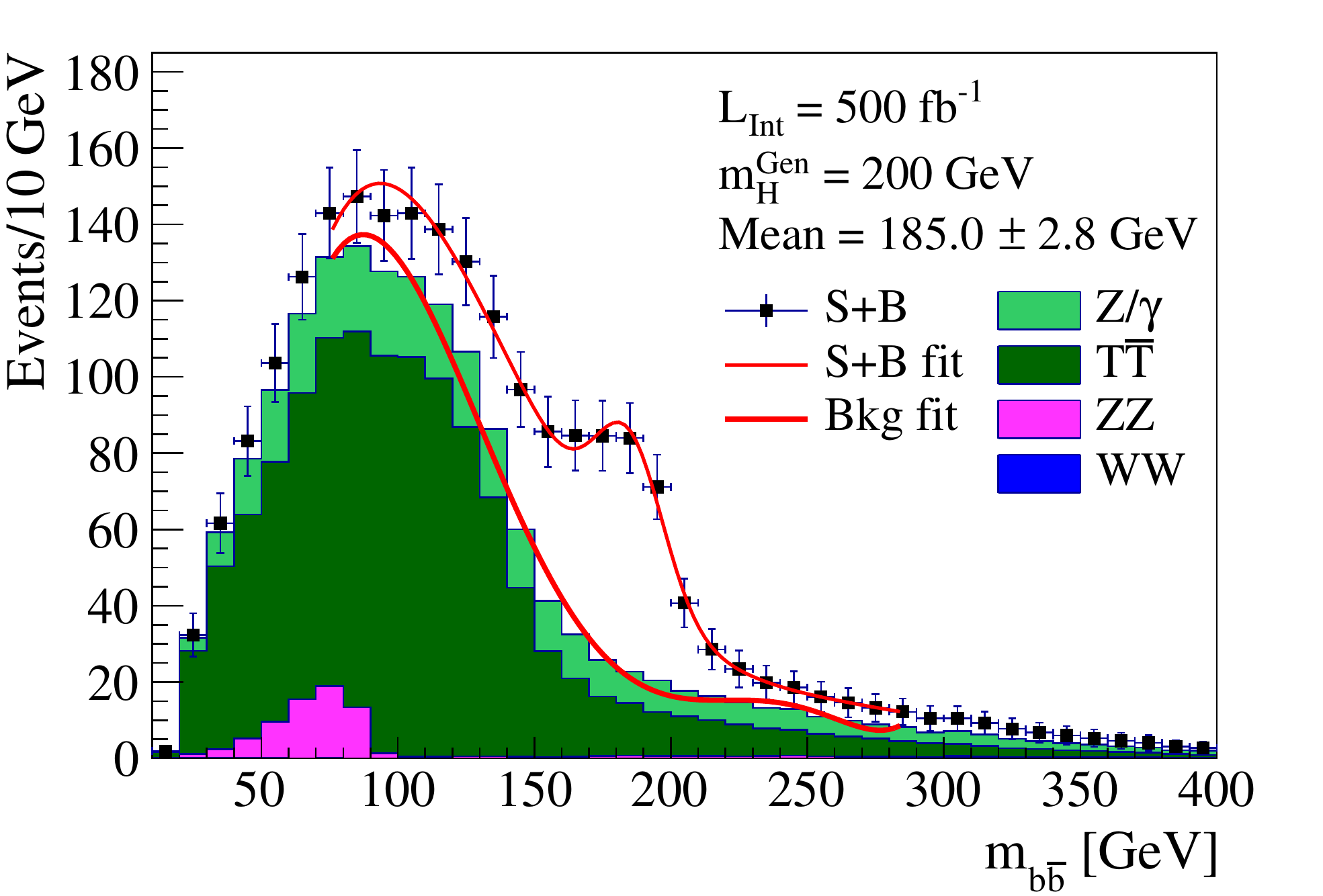}
    \caption{}
    \label{HBP3}
    \end{subfigure}
    \begin{subfigure}[b]{0.49\textwidth}
    \centering
    \includegraphics[width=\textwidth]{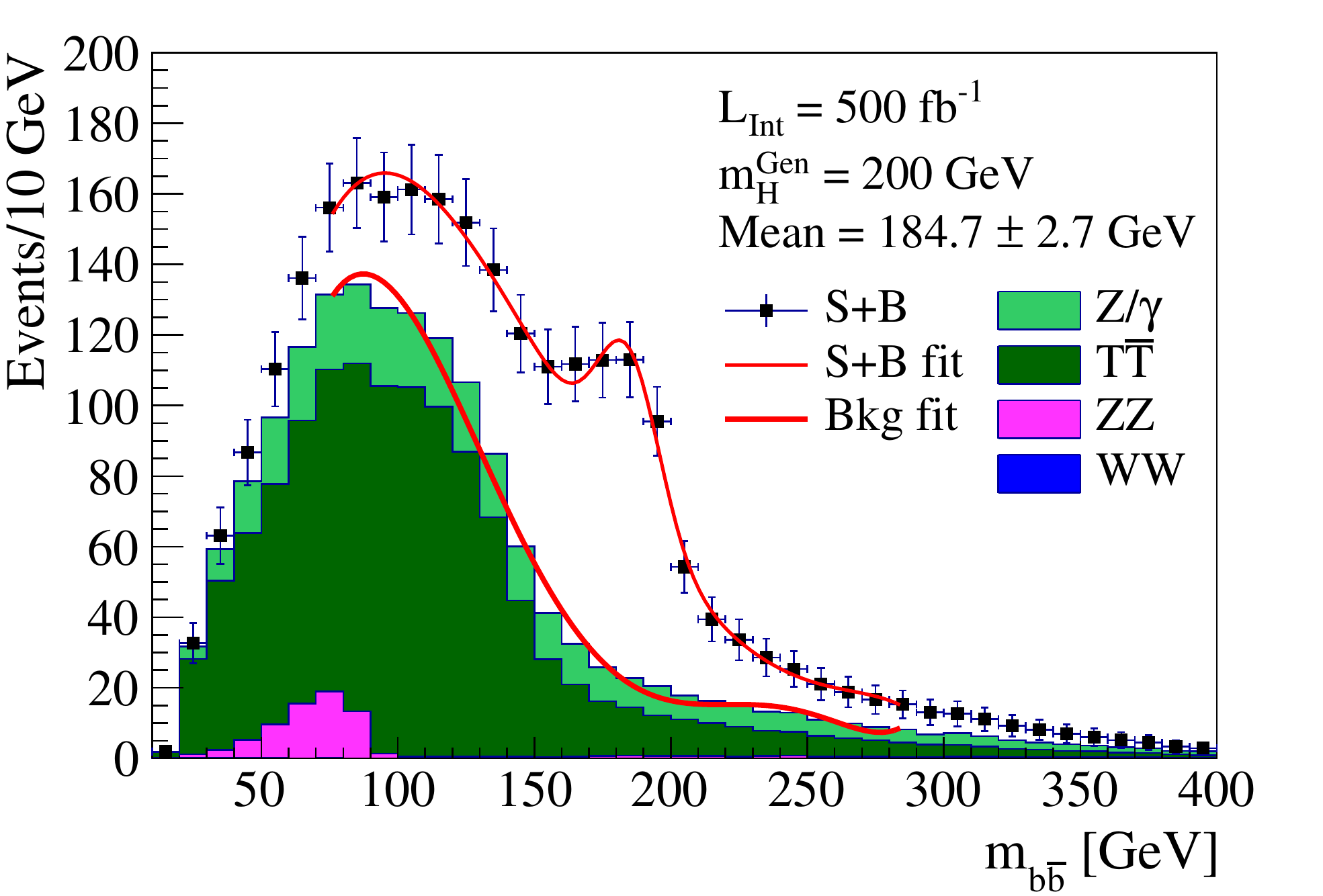}
    \caption{}
    \label{HBP4}
    \end{subfigure}
    \begin{subfigure}[b]{0.49\textwidth}
    \centering
    \includegraphics[width=\textwidth]{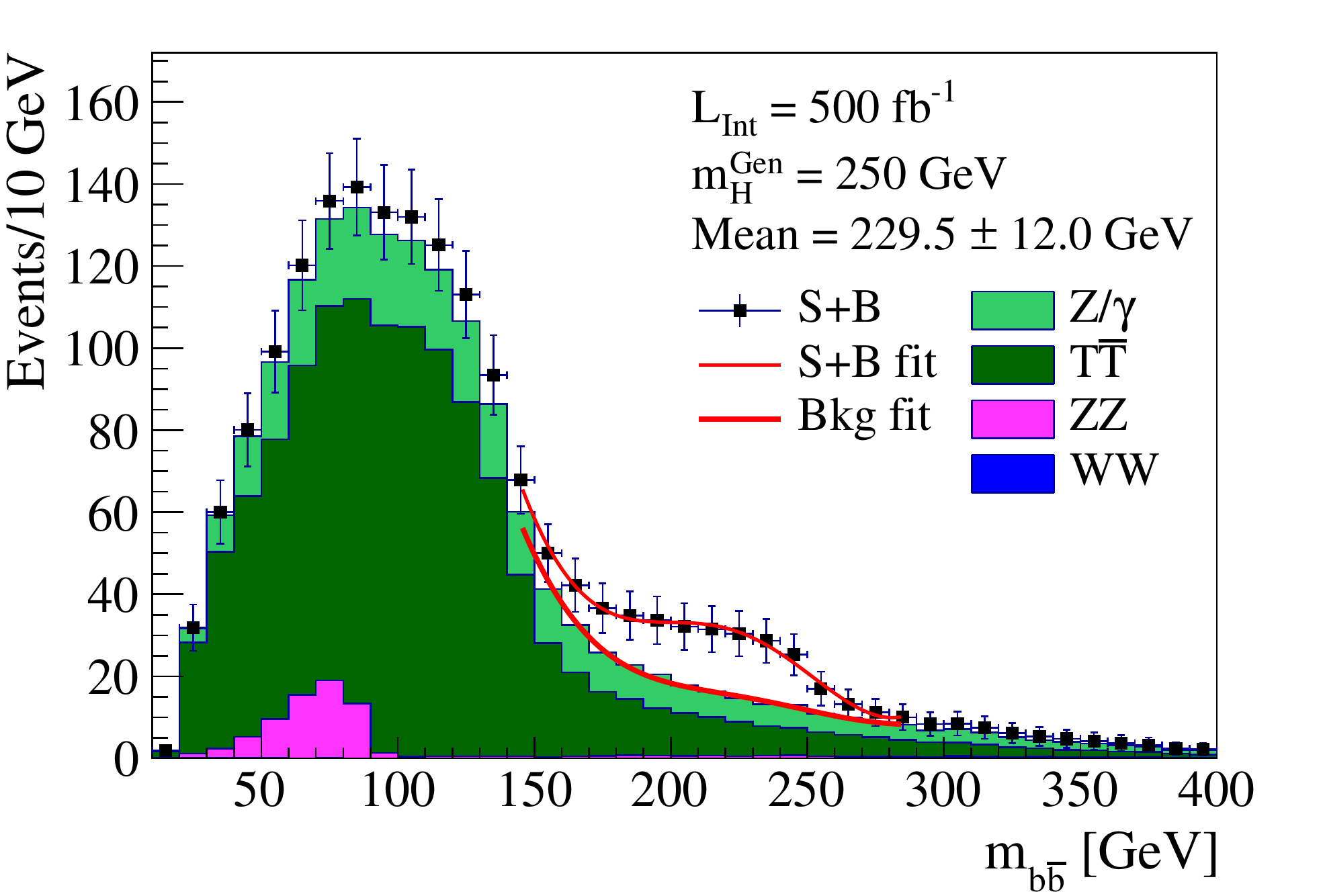}
    \caption{}
    \label{HBP5}
    \end{subfigure}
\begin{subfigure}[b]{0.49\textwidth}
    \centering
    \includegraphics[width=\textwidth]{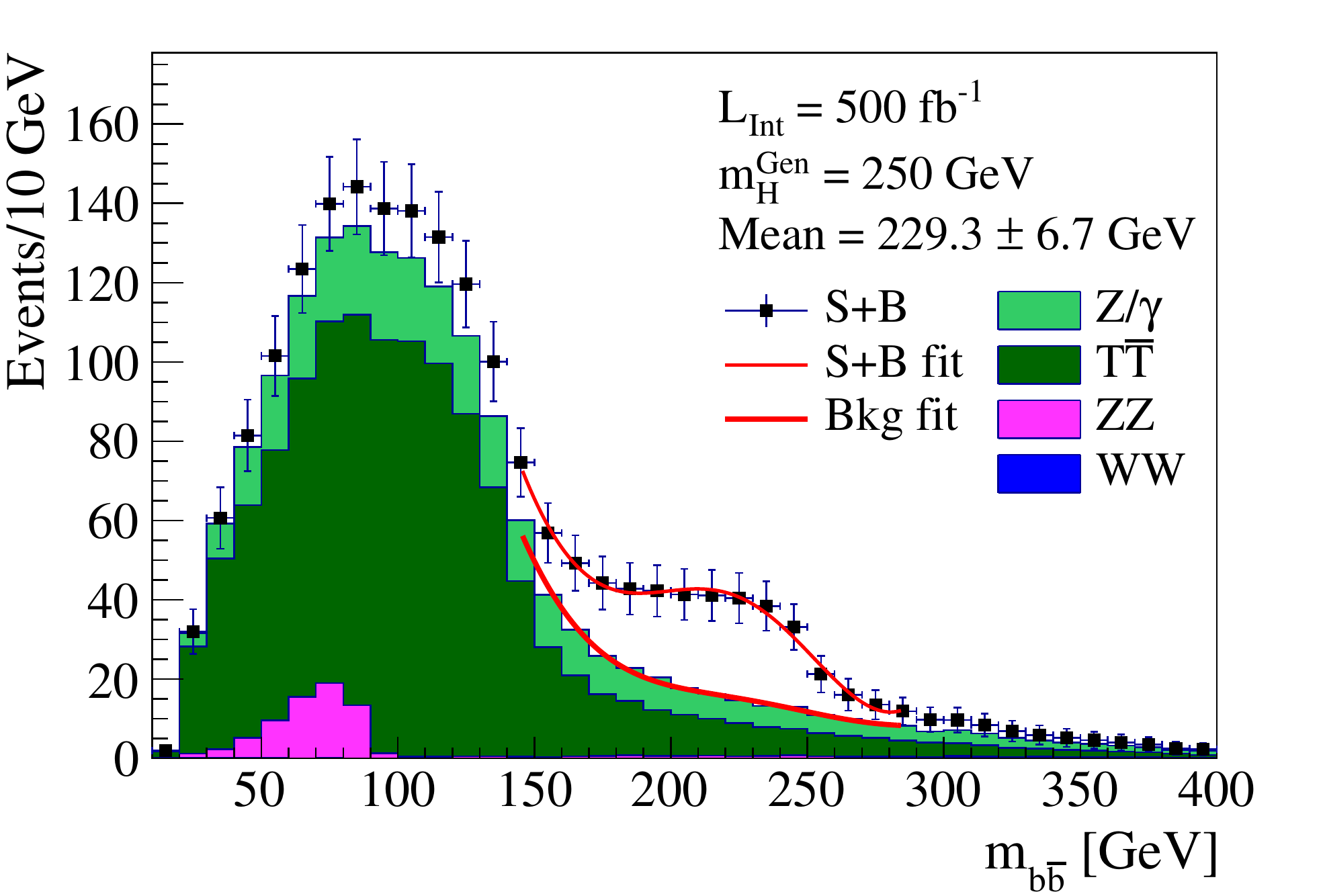}
    \caption{}
    \label{HBP6}
    \end{subfigure}
\caption{Candidate mass distributions of the a,b,c,d,e,f) $H$ and g,h,i,j,k,l) $A$ Higgs bosons with corresponding fitting results and errors in different scenarios at $\sqrt s=1000$ GeV.}
\label{1000fit}
\end{figure*}
\begin{figure*}[h]\ContinuedFloat
    \begin{subfigure}[b]{0.49\textwidth}
    \centering
    \includegraphics[width=\textwidth]{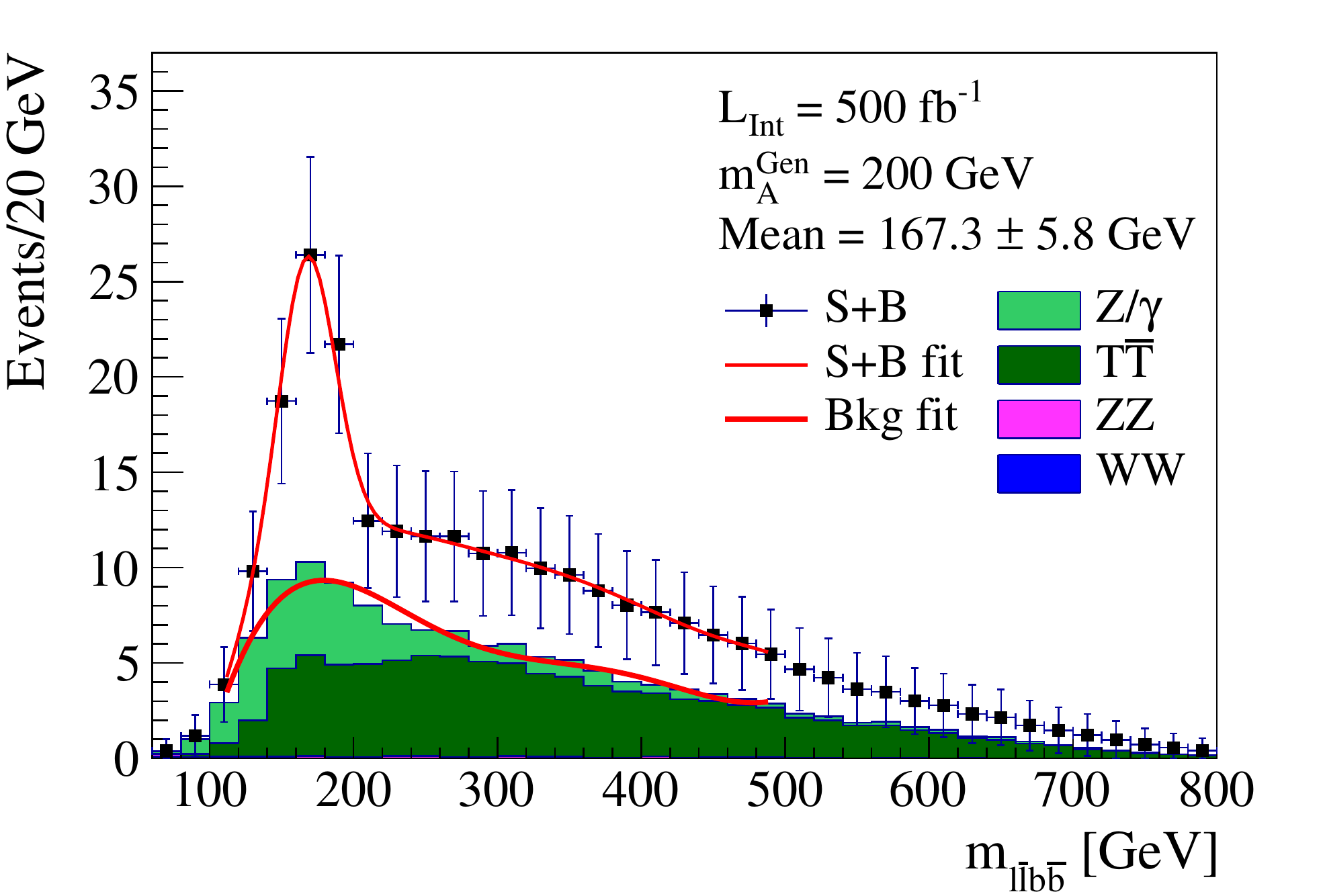}
    \caption{}
    \label{ABP1}
    \end{subfigure}
    \begin{subfigure}[b]{0.49\textwidth}
    \centering
    \includegraphics[width=\textwidth]{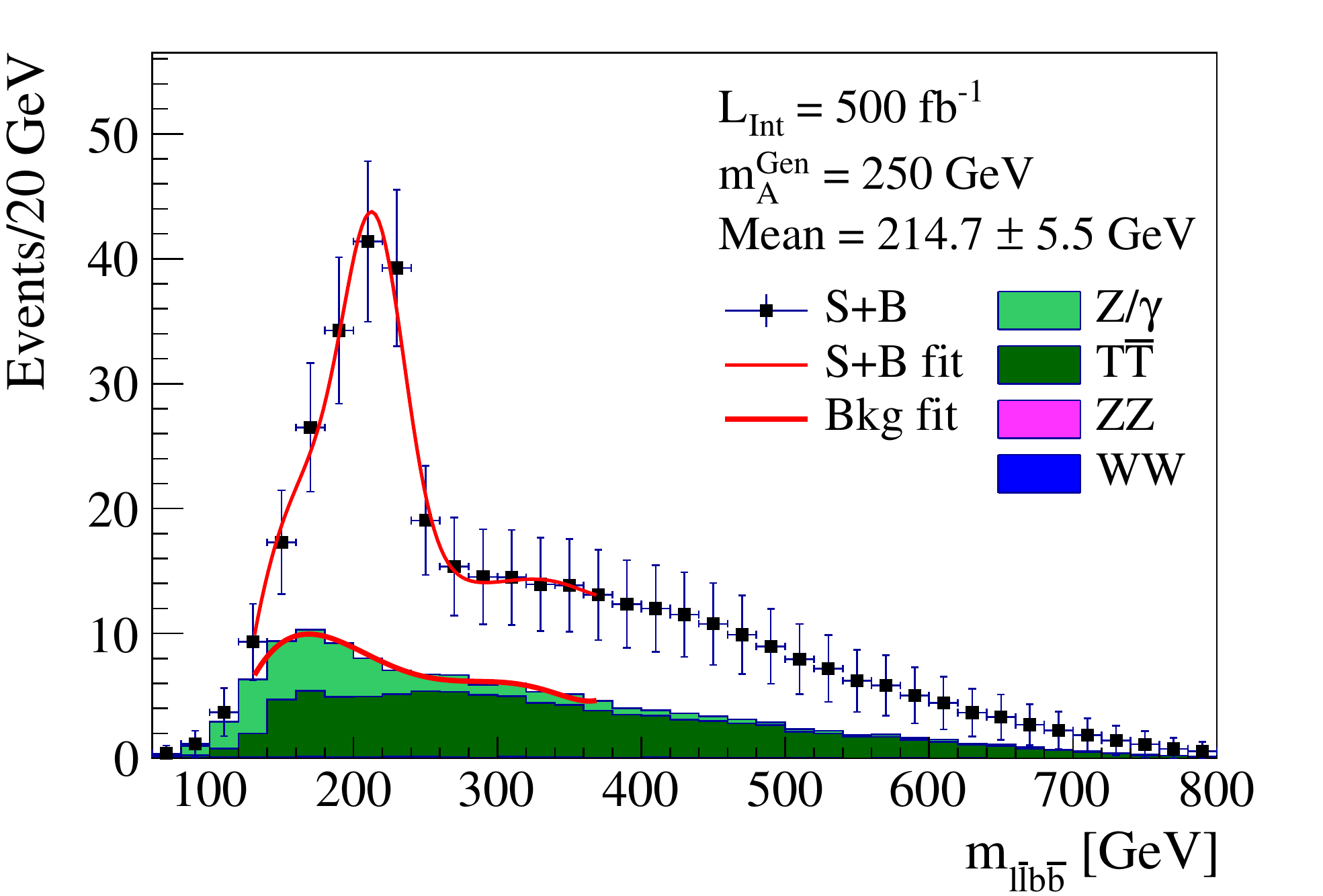}
    \caption{}
    \label{ABP2}
    \end{subfigure}
    \begin{subfigure}[b]{0.49\textwidth}
    \centering
    \includegraphics[width=\textwidth]{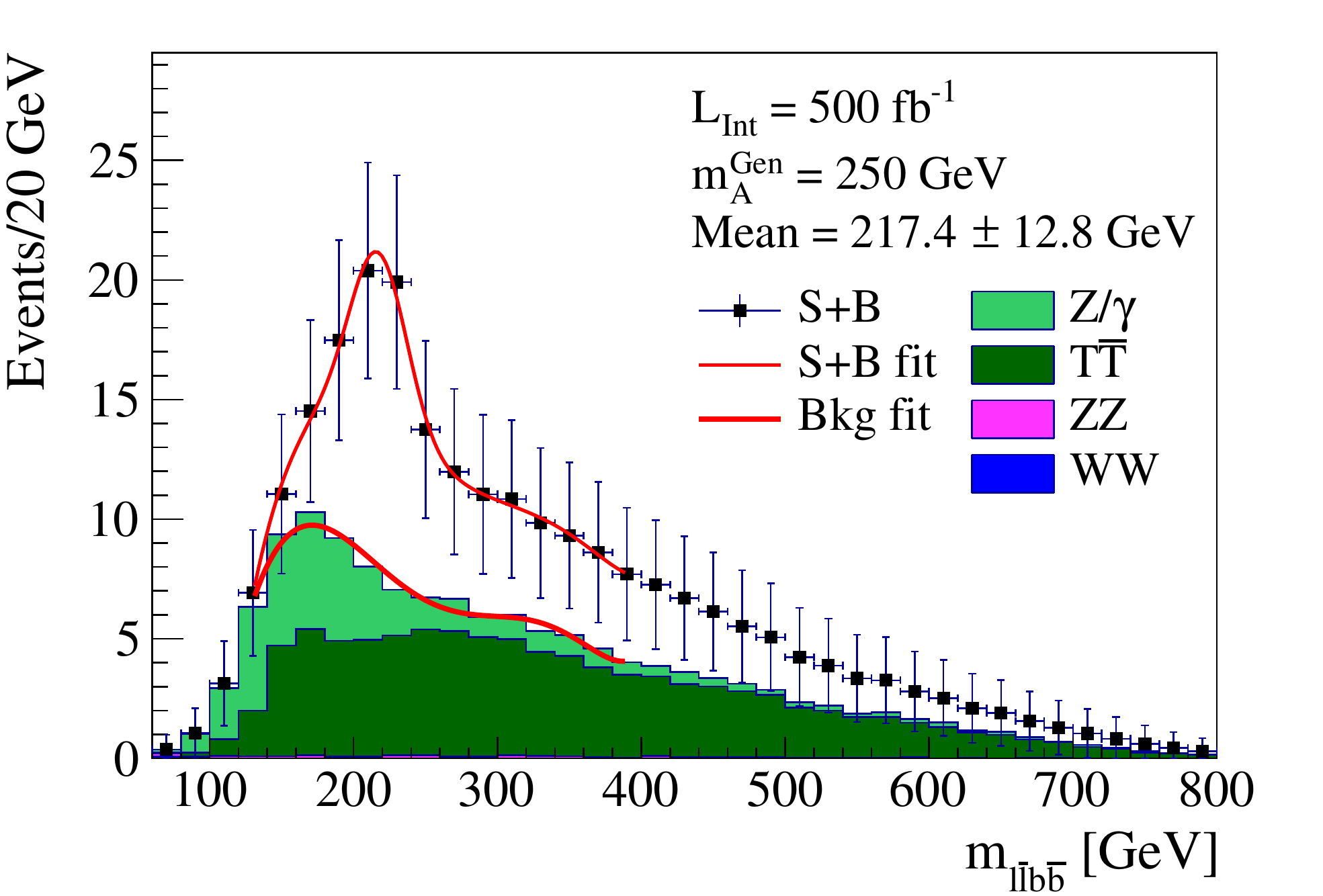} 
    \caption{}   
    \label{ABP3}
    \end{subfigure}
    \begin{subfigure}[b]{0.49\textwidth}
    \centering
    \includegraphics[width=\textwidth]{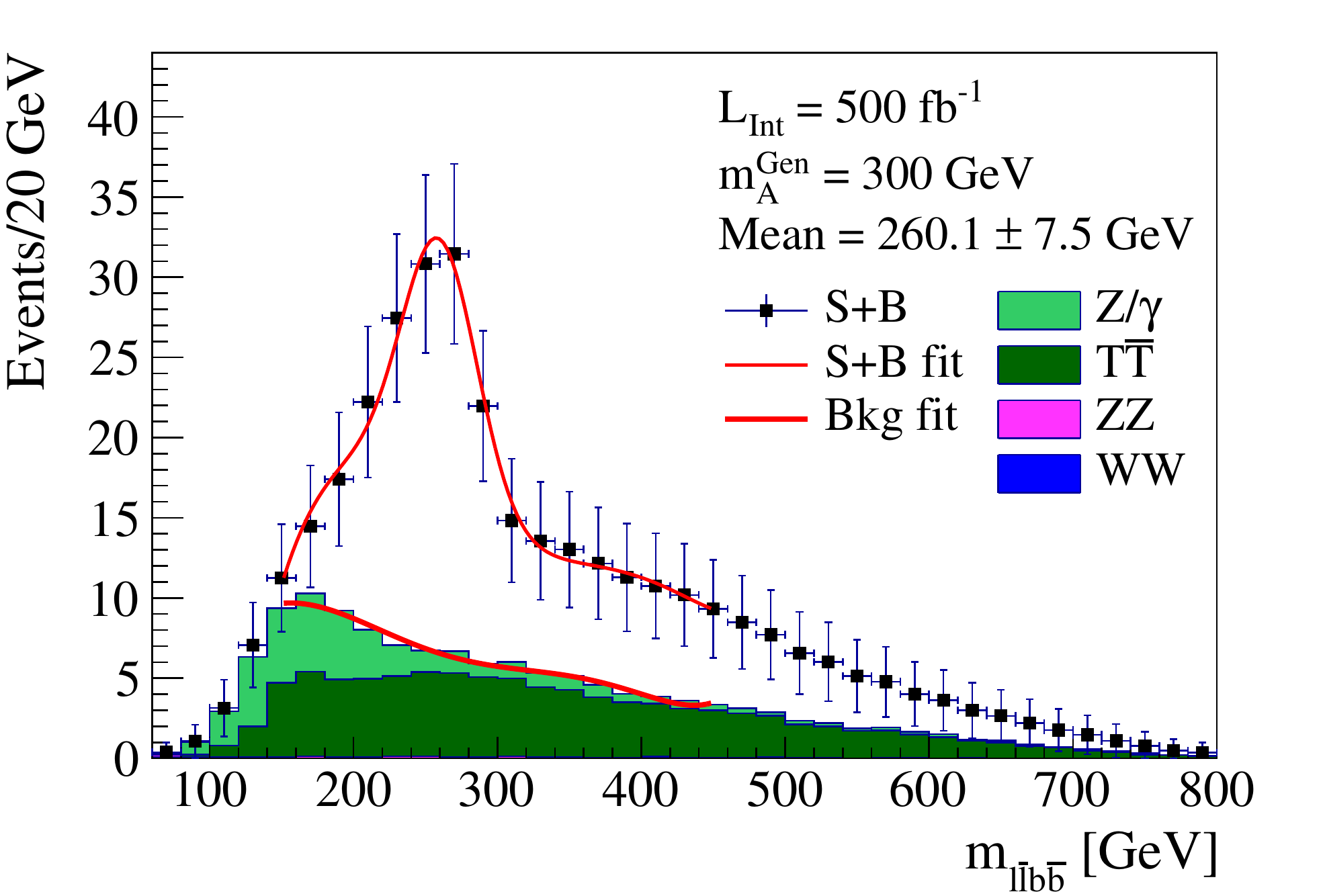}
    \caption{}
    \label{ABP4}
    \end{subfigure}
    \begin{subfigure}[b]{0.49\textwidth}
    \centering
    \includegraphics[width=\textwidth]{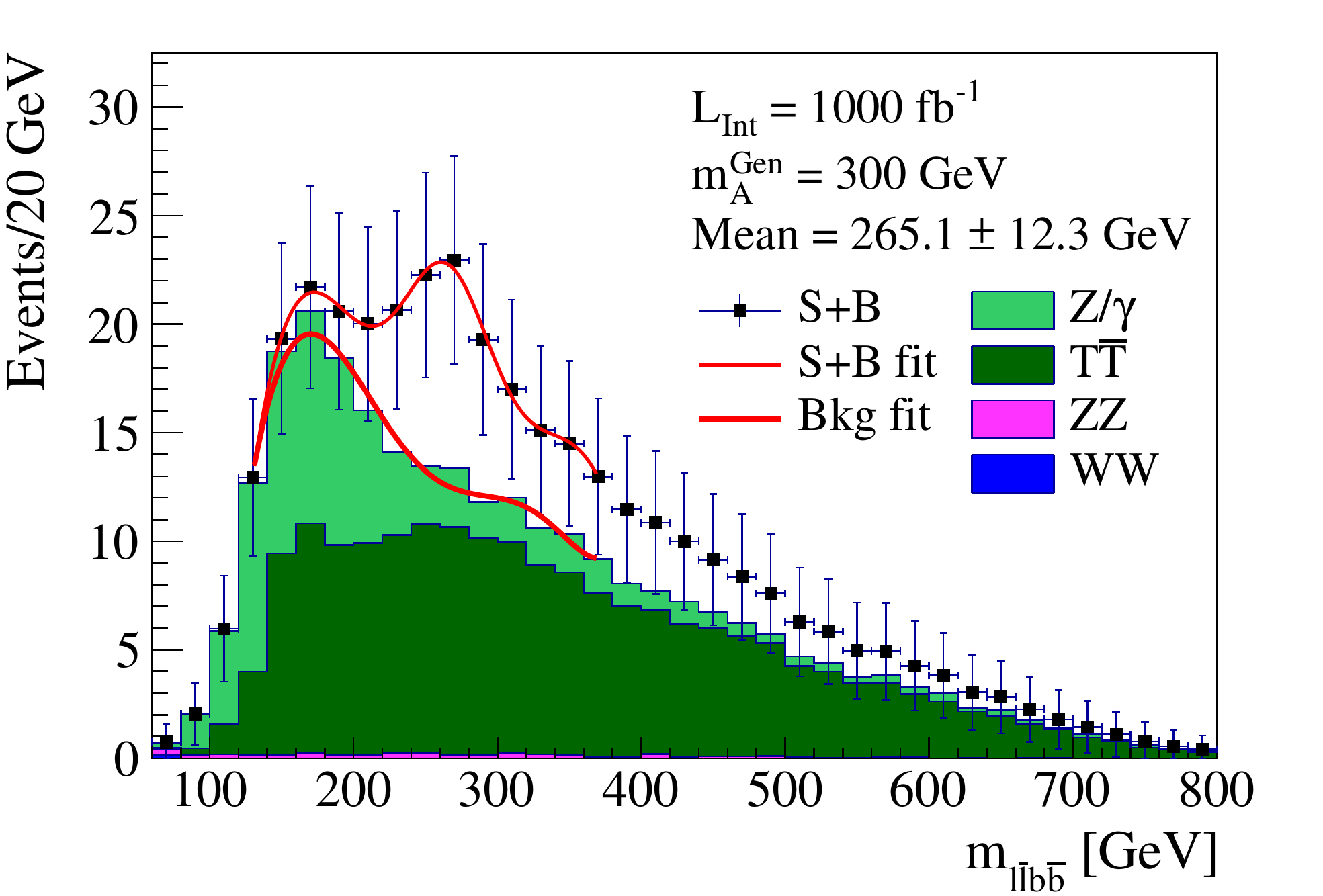}
    \caption{}
    \label{ABP5}
    \end{subfigure}
\begin{subfigure}[b]{0.49\textwidth}
    \centering
    \includegraphics[width=\textwidth]{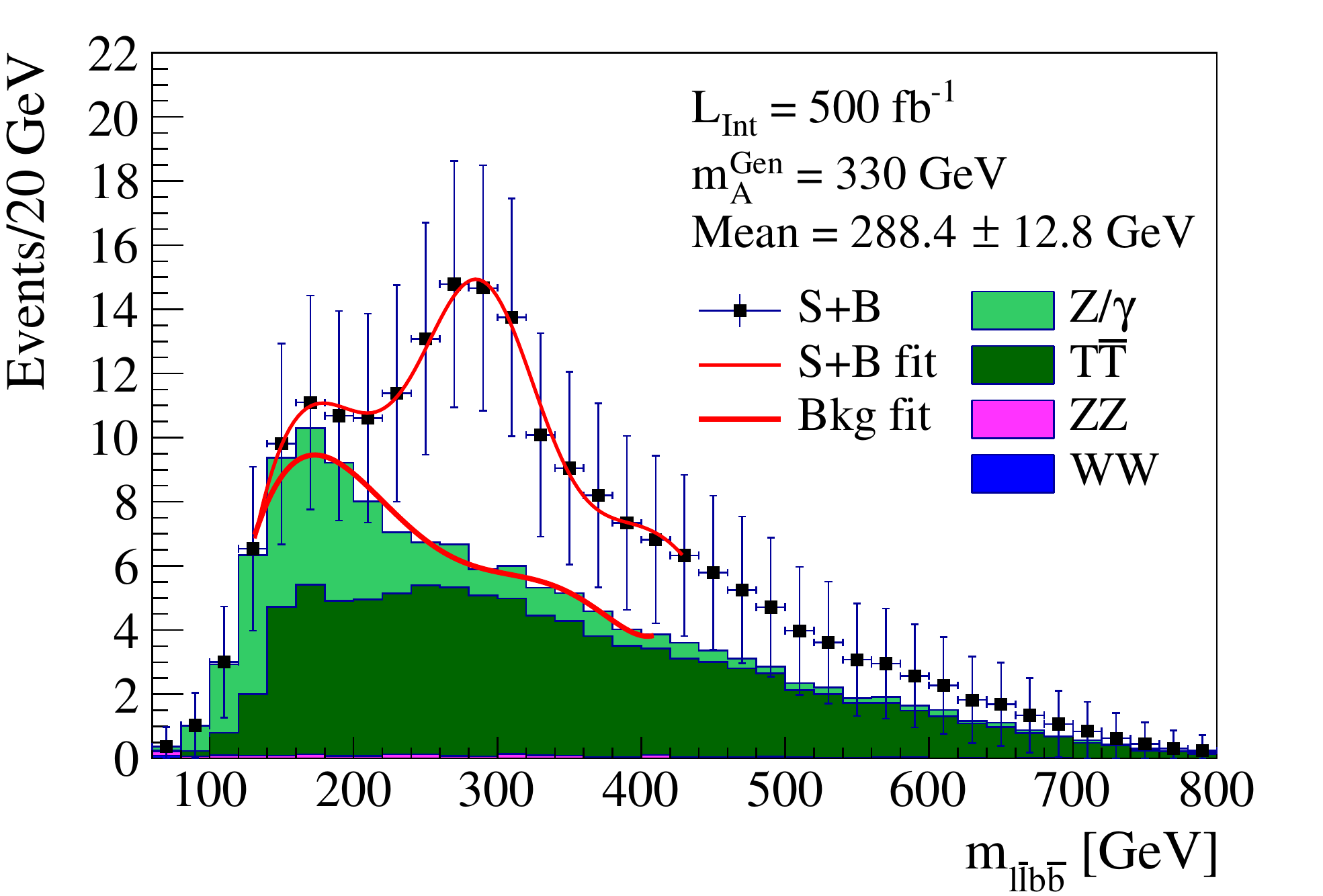}
    \caption{}
    \label{ABP6}
    \end{subfigure}
\caption{Candidate mass distributions of the a,b,c,d,e,f) $H$ and g,h,i,j,k,l) $A$ Higgs bosons with corresponding fitting results and errors in different scenarios at $\sqrt s=1000$ GeV.}
  \label{1000fit}
\end{figure*}
Computing the invariant mass of the identified $b\bar{b}$ and $jjb\bar{b}$ combinations for events surviving the selection cuts, Higgs candidate mass distributions of Figs. \ref{500fit} and \ref{1000fit} are obtained for $\sqrt s= 500$ and $1000$ GeV respectively. Distributions corresponding to different scenarios are shown separately for each one of the Higgs bosons. Signal and different background contributions are also shown separately. As seen, the $t\bar{t}$ process has the most contribution among the background processes, and is, however, well under control. In all of the distributions, signal contribution is seen as a significant excess of data on top of the total background. The distributions are normalized based on $L\times\sigma\times\epsilon$, where $L$ is the integrated luminosity, $\sigma$ is the cross section and $\epsilon$ is the selection efficiency. The integrated luminosity is set to 500 $fb^{-1}$ for all of the distributions except for the distribution of Fig. \ref{ABP5} which uses the integrated luminosity of 1000 $fb^{-1}$. Signal cross sections are obtained by multiplying the total cross sections provided in Tab. \ref{BPs} by corresponding branching ratios of the decay modes $A\rightarrow ZH$, $Z\rightarrow jj$ and $H\rightarrow b\bar{b}$. Background cross sections are taken from Tab. \ref{bgXsec}. Selection efficiencies for $A$ mass distributions are taken from Tab. \ref{eff} and efficiencies corresponding to $H$ mass distributions are obtained by computing the total number of signal reconstructed $H$ Higgs bosons divided by twice the number of simulated signal events.

Proper fit functions are fitted to the total background (B) and signal+total background (S+B) distributions of Figs. \ref{500fit} and \ref{1000fit} and results are shown with associated error bars. Fitted curves show significant peaks near the generated Higgs masses. Fitting is performed by ROOT 5.34 \cite{root2}. The fit function used for B distributions is a polynomial function and the fit function used for S+B distributions is the combination of a polynomial and a Gaussian function. The Gaussian function covers the signal and the polynomial covers the total background. First, the polynomial function is fitted to the total background, and then the parameters of the fitted polynomial are used as input for S+B fit. The ``Mean'' parameter is one of the Gaussian fit function parameters and provides the center of the signal peak. The values obtained for the ``Mean'' parameter are considered as the reconstructed masses ($m_{\,Rec.}$) of the Higgs bosons and are provided in Tab. \ref{recmass}.
\begin{table}[h]
\begin{subtable}[b]{.48\textwidth}
\normalsize
\fontsize{11}{7.2} 
    \begin{center}
         \begin{tabular}{ >{\centering\arraybackslash}m{.18in} >{\centering\arraybackslash}m{.27in}  >{\centering\arraybackslash}m{.5in}  >{\centering\arraybackslash}m{.9in} >{\centering\arraybackslash}m{.9in} >{\centering\arraybackslash}m{0in} } 
\multicolumn{6}{ c }{\,\,\,\,\,\,\,\,\,\,\,\,\,\,\,\,\,\,\,\,\,\,\,\,\,\,\,\,\,\,\,\,\,\,\,\,\,\,\,\,\,\,$\sqrt s=500$ GeV\,\,\,\,\,\,\,\,\,\,\,\,\,\,\,\,\,\,\,\,\,\,\,\,\,\,\,\,\,\,\,\,\,\,\,\,\,\,\,\,\,\,\,\,} \\ \Xhline{3\arrayrulewidth} 
  &  & $m_{\,Gen.}$ & $m_{\, Rec.}$ & $m_{\,\, Corr. \,\, rec.}$ \parbox{0pt}{\rule{0pt}{1ex+\baselineskip}}\\ \Xhline{3\arrayrulewidth}
  \multirow{2}{*}[-3.1pt]{\textbf{H}} &   \cellcolor{blizzardblue}{BP1} & 150 & 127.3 $\pm$ 2.3 & 150.2$\pm$5.0 \parbox{0pt}{\rule{0pt}{1ex+\baselineskip}}\\ 
  &   \cellcolor{blizzardblue}{BP2} & 150 & 127.0$\pm$3.0 & 149.9$\pm$5.7 \parbox{0pt}{\rule{0pt}{1ex+\baselineskip}}\\ \Xhline{2\arrayrulewidth}
 \multirow{2}{*}[-3.1pt]{\textbf{A}} &   \cellcolor{blizzardblue}{BP1} & 200 & 170.8$\pm$7.1 & 200.4$\pm$14.3 \parbox{0pt}{\rule{0pt}{1ex+\baselineskip}}\\ 
  &   \cellcolor{blizzardblue}{BP2} & 250 & 210.1$\pm$7.2  & 239.7$\pm$14.4 \parbox{0pt}{\rule{0pt}{1ex+\baselineskip}}\\ \Xhline{3\arrayrulewidth}
        \end{tabular}
\caption{}  
 \label{recmass500}   
  \end{center}  
\end{subtable} 
\newline\newline
\begin{subtable}[b]{.48\textwidth}
\normalsize
\fontsize{11}{7.2} 
    \begin{center}
         \begin{tabular}{ >{\centering\arraybackslash}m{.18in} >{\centering\arraybackslash}m{.27in}  >{\centering\arraybackslash}m{.5in}  >{\centering\arraybackslash}m{.9in} >{\centering\arraybackslash}m{.9in} >{\centering\arraybackslash}m{0in} } 
\multicolumn{6}{ c }{\,\,\,\,\,\,\,\,\,\,\,\,\,\,\,\,\,\,\,\,\,\,\,\,\,\,\,\,\,\,\,\,\,\,\,\,\,\,\,\,$\sqrt s=1000$ GeV\,\,\,\,\,\,\,\,\,\,\,\,\,\,\,\,\,\,\,\,\,\,\,\,\,\,\,\,\,\,\,\,\,\,\,\,\,\,\,\,\,\,} \\ \Xhline{3\arrayrulewidth} 
  &  & $m_{\,Gen.}$ & $m_{\, Rec.}$ & $m_{\,\, Corr. \,\, rec.}$ \parbox{0pt}{\rule{0pt}{1ex+\baselineskip}}\\ \Xhline{3\arrayrulewidth}
   &   \cellcolor{blizzardblue}{BP1} & 150 & 132.4$\pm$1.8 & 150.3$\pm$6.4 \parbox{0pt}{\rule{0pt}{1ex+\baselineskip}}\\ 
  &   \cellcolor{blizzardblue}{BP2} & 150 & 132.0$\pm$1.7 & 149.9$\pm$6.3 \parbox{0pt}{\rule{0pt}{1ex+\baselineskip}}\\  
\multirow{2}{*}[-3.15pt]{\textbf{H}} &   \cellcolor{blizzardblue}{BP3} & 200 & 185.0$\pm$2.8 & 202.9$\pm$7.4 \parbox{0pt}{\rule{0pt}{1ex+\baselineskip}}\\
&   \cellcolor{blizzardblue}{BP4} & 200 & 184.7$\pm$2.7 & 202.6$\pm$7.3 \parbox{0pt}{\rule{0pt}{1ex+\baselineskip}}\\ 
&   \cellcolor{blizzardblue}{BP5} & 250 & 229.5$\pm$12.0 & 247.4$\pm$16.6 \parbox{0pt}{\rule{0pt}{1ex+\baselineskip}}\\ 
&   \cellcolor{blizzardblue}{BP6} & 250 & 229.3$\pm$6.7 & 247.2$\pm$11.3 \parbox{0pt}{\rule{0pt}{1ex+\baselineskip}}\\ \Xhline{2\arrayrulewidth}
 &   \cellcolor{blizzardblue}{BP1} & 200 & 167.3$\pm$5.8 & 203.5$\pm$15.3 \parbox{0pt}{\rule{0pt}{1ex+\baselineskip}}\\ 
  &   \cellcolor{blizzardblue}{BP2} & 250 & 214.7$\pm$5.5  & 250.9$\pm$15.0 \parbox{0pt}{\rule{0pt}{1ex+\baselineskip}}\\ 
 \multirow{2}{*}[-3.15pt]{\textbf{A}} & \cellcolor{blizzardblue}{BP3} & 250 & 217.4$\pm$12.8 & 253.6$\pm$22.3 \parbox{0pt}{\rule{0pt}{1ex+\baselineskip}}\\
&   \cellcolor{blizzardblue}{BP4} & 300 & 260.1$\pm$7.5 & 296.3$\pm$17.0 \parbox{0pt}{\rule{0pt}{1ex+\baselineskip}}\\
&   \cellcolor{blizzardblue}{BP5} & 300 & 265.1$\pm$12.3 & 301.3$\pm$21.8 \parbox{0pt}{\rule{0pt}{1ex+\baselineskip}}\\ 
&   \cellcolor{blizzardblue}{BP6} & 330 & 288.4$\pm$12.8 & 324.6$\pm$22.3 \parbox{0pt}{\rule{0pt}{1ex+\baselineskip}}\\ \Xhline{3\arrayrulewidth}
        \end{tabular}
\caption{} 
 \label{recmass1000}
  \end{center} 
\end{subtable}  
\caption{Generated mass ($m_{\,Gen.}$), reconstructed mass ($m_{\,Rec.}$) and corrected reconstructed mass ($m_{\,\,Corr.\,\, rec.}$) of the Higgs bosons $H$ and $A$ with associated uncertainties at $\sqrt s=$ a) 500 and b) 1000 GeV. The mass values are provided in GeV unit.}
\label{recmass} 
\end{table} 
The generated masses ($m_{\,Gen.}$) are also shown for comparison. According to Tab. \ref{recmass}, a difference is seen between the generated and reconstructed masses which can be explained by the uncertainties arising from jet clustering algorithm and jet mis-identification, jet mis-tag rate, fitting method and choice of the fit function, errors in energy and momentum of the particles, etc. Optimization of the jet clustering algorithm, $b$-tagging algorithm and fitting method may reduce the errors. However, since such optimizations lie beyond the scope of this paper, a simple off-set correction is applied to reduce the errors in this study. The off-set correction is applied as follows. According to Tab. \ref{recmass500}, on average, the reconstructed masses of the $H$ and $A$ Higgs bosons are 22.85 and 29.55 GeV smaller than the corresponding generated masses respectively. Calculating the same values for Tab. \ref{recmass1000}, we obtain 17.85 and 36.17 GeV for the Higgs bosons $H$ and $A$ respectively. To reduce the errors, reconstructed $H$ and $A$ Higgs masses are increased by the same values and the results are provided in Tab. \ref{recmass} as corrected reconstructed masses ($m_{\,\,Corr.\,\, rec.}$). According to the results, corrected reconstructed masses are obtained with few GeVs difference from the generated masses in the considered scenarios and it is concluded that the masses of the Higgs bosons $H$ and $A$ are measurable in all of the assumed scenarios.

\section{Signal significance}   
To assess the observability of the Higgs bosons in the considered scenarios, signal significance is computed for each of the candidate mass distributions of Figs. \ref{500fit} and \ref{1000fit} by counting the number of signal and background candidate masses in the whole mass range. 
\begin{table*}[!htbp]
\begin{subtable}[b]{.33\textwidth}
\normalsize
\fontsize{11}{7.2} 
    \begin{center}   
         \begin{tabular}{ >{\centering\arraybackslash}m{.16in} >{\centering\arraybackslash}m{.9in}  >{\centering\arraybackslash}m{.35in}  >{\centering\arraybackslash}m{.36in} >{\centering\arraybackslash}m{.35in}  }  
\multicolumn{5}{ c }{\,\,\,\,\,\,\,\,\,\,\,\,\,\,\,\,\,\,\,\,\,$\sqrt s=500$ GeV\,\,\,\,\,\,\,\,\,\,\,\,\,\,\,\,\,\,\,\,\,\,\,} \\ \Xhline{3\arrayrulewidth}
  &  & {BP1} & {BP2} \parbox{0pt}{\rule{0pt}{1ex+\baselineskip}}\\ \Xhline{3\arrayrulewidth} 
  \multirow{9}{*}[4.8pt]{\textbf{H}} 
 &  \cellcolor{blizzardblue}{$\epsilon_{\, Total}$} & 0.21 & 0.25 \parbox{0pt}{\rule{0pt}{1ex+\baselineskip}}\\ 
  &  \cellcolor{blizzardblue}{$S$} & 790.6 & 707.1  \parbox{0pt}{\rule{0pt}{1ex+\baselineskip}}\\ 
  &  \cellcolor{blizzardblue}{$B$} & \multicolumn{2}{c}{1397.2} \parbox{0pt}{\rule{0pt}{1ex+\baselineskip}}\\  
&\cellcolor{blizzardblue}{$S/B$} & 0.57 & 0.51 \parbox{0pt}{\rule{0pt}{1ex+\baselineskip}}\\  
&\cellcolor{blizzardblue}{$S/\sqrt{B}$} & 21.2 & 18.9  \parbox{0pt}{\rule{0pt}{1ex+\baselineskip}}\\ 
&  \cellcolor{blizzardblue}{$\mathcal{L}_{\, Int.}$ [$fb^{-1}$]} &  \multicolumn{2}{c}{500} \parbox{0pt}{\rule{0pt}{1ex+\baselineskip}}\\ 
\Xhline{3\arrayrulewidth} 
 \multirow{9}{*}[4.8pt]{\textbf{A}} 
  &  \cellcolor{blizzardblue}{$\epsilon_{\, Total}$} & 0.15 & 0.21  \parbox{0pt}{\rule{0pt}{1ex+\baselineskip}}\\ 
  &  \cellcolor{blizzardblue}{$S$} & 283.9 & 294.6   \parbox{0pt}{\rule{0pt}{1ex+\baselineskip}}\\ 
  &  \cellcolor{blizzardblue}{$B$} & \multicolumn{2}{c}{459.6} \parbox{0pt}{\rule{0pt}{1ex+\baselineskip}}\\  
&\cellcolor{blizzardblue}{$S/B$} & 0.62 & 0.64  \parbox{0pt}{\rule{0pt}{1ex+\baselineskip}}\\ 
&\cellcolor{blizzardblue}{$S/\sqrt{B}$} & 13.2 & 13.7   \parbox{0pt}{\rule{0pt}{1ex+\baselineskip}}\\ 
&  \cellcolor{blizzardblue}{$\mathcal{L}_{\, Int.}$ [$fb^{-1}$]}  &  \multicolumn{2}{c}{500} \parbox{0pt}{\rule{0pt}{1ex+\baselineskip}}\\
\Xhline{3\arrayrulewidth} 
        \end{tabular}  
\caption{}   
 \label{sign500}  
  \end{center}  
\end{subtable} 
\begin{subtable}[b]{.58\textwidth}
\normalsize
\fontsize{11}{7.2} 
    \begin{center} 
         \begin{tabular}{ >{\centering\arraybackslash}m{.16in} >{\centering\arraybackslash}m{.9in}  >{\centering\arraybackslash}m{.35in}  >{\centering\arraybackslash}m{.35in} >{\centering\arraybackslash}m{.39in}  >{\centering\arraybackslash}m{.35in} >{\centering\arraybackslash}m{.45in} >{\centering\arraybackslash}m{.36in} >{\centering\arraybackslash}m{.35in}}  
\multicolumn{9}{ c }{\,\,\,\,\,\,\,\,\,\,\,\,\,\,\,\,\,\,\,\,\,\,\,\,\,\,\,\,\,\,\,\,\,\,\,\,\,\,\,\,\,\,\,\,\,\,\,\,\,\,\,\,\,\,\,\,\,$\sqrt s=1000$ GeV\,\,\,\,\,\,\,\,\,\,\,\,\,\,\,\,\,\,\,\,\,\,\,\,\,\,\,\,\,\,\,\,\,\,\,\,\,\,\,\,\,\,\,\,\,\,\,\,\,\,\,\,\,\,\,\,\,\,\,} \\ \Xhline{3\arrayrulewidth}
  &  & {BP1} & {BP2} & {BP3} & {BP4} & {BP5} & {BP6} \parbox{0pt}{\rule{0pt}{1ex+\baselineskip}}\\ \Xhline{3\arrayrulewidth} 
  \multirow{9}{*}[4.8pt]{\textbf{H}}  
&  \cellcolor{blizzardblue}{$\epsilon_{\, Total}$} & 0.29 & 0.35 & 0.40 & 0.44 & 0.43 & 0.46 \parbox{0pt}{\rule{0pt}{1ex+\baselineskip}}\\ 
  &  \cellcolor{blizzardblue}{$S$} & 799.0 & 1306.9 & 555.5 & 942.3 & 206.9 & 364.0 \parbox{0pt}{\rule{0pt}{1ex+\baselineskip}}\\ 
  &  \cellcolor{blizzardblue}{$B$} & \multicolumn{6}{c}{1586.1} \parbox{0pt}{\rule{0pt}{1ex+\baselineskip}}\\  
&\cellcolor{blizzardblue}{$S/B$} & 0.50 & 0.82 & 0.35 & 0.59 & 0.13 & 0.23 \parbox{0pt}{\rule{0pt}{1ex+\baselineskip}}\\  
&\cellcolor{blizzardblue}{$S/\sqrt{B}$} & 20.1 & 32.8 & 13.9 & 23.7 & 5.2 & 9.1 \parbox{0pt}{\rule{0pt}{1ex+\baselineskip}}\\  
&  \cellcolor{blizzardblue}{$\mathcal{L}_{\, Int.}$ [$fb^{-1}$]} &  \multicolumn{6}{c}{500} \parbox{0pt}{\rule{0pt}{1ex+\baselineskip}}\\ 
\Xhline{3\arrayrulewidth} 
 \multirow{9}{*}[4.8pt]{\textbf{A}}  &  \cellcolor{blizzardblue}{$\epsilon_{\, Total}$} & 0.09 & 0.14 & 0.15 & 0.20 & 0.18 & 0.20 \parbox{0pt}{\rule{0pt}{1ex+\baselineskip}}\\ 
  &  \cellcolor{blizzardblue}{$S$} & 121.2 & 265.0 & 106.0 & 211.3 & 84.2 & 78.4 \parbox{0pt}{\rule{0pt}{1ex+\baselineskip}}\\ 
  &  \cellcolor{blizzardblue}{$B$} & \multicolumn{4}{c}{133.0} & 266.1 & 133.0 \parbox{0pt}{\rule{0pt}{1ex+\baselineskip}}\\  
&\cellcolor{blizzardblue}{$S/B$} & 0.91 & 1.99 & 0.80 & 1.59 & 0.32 & 0.59 \parbox{0pt}{\rule{0pt}{1ex+\baselineskip}}\\ 
&\cellcolor{blizzardblue}{$S/\sqrt{B}$} & 10.5 & 23.0 & 9.2 & 18.3 & 5.2 & 6.8 \parbox{0pt}{\rule{0pt}{1ex+\baselineskip}}\\ 
&  \cellcolor{blizzardblue}{$\mathcal{L}_{\, Int.}$ [$fb^{-1}]$} &  \multicolumn{4}{c}{500} & 1000 & 500 \parbox{0pt}{\rule{0pt}{1ex+\baselineskip}}\\
\Xhline{3\arrayrulewidth}
        \end{tabular}
\caption{} 
 \label{sign1000}   
  \end{center}
\end{subtable} 
\caption{Total signal selection efficiency ($\epsilon_{\, Total}$), number of signal (S) and background (B) candidate masses, signal to background ratio, signal significance and the assumed integrated luminosity in different scenarios at the center-of-mass energies of a) 500 and b) 1000 GeV.} 
 \label{sign}
\end{table*}
Computation of the signal significance is based on the integrated luminosity of 500 $fb^{-1}$ for all of the distributions except for the distribution of Fig. \ref{ABP5} which uses the integrated luminosity of 1000 $fb^{-1}$. Computation results, namely total signal selection efficiency, number of signal (S) and background (B) candidate masses, signal to background ratio and signal significance at the center-of-mass energies of 500 and 1000 GeV are provided in Tab. \ref{sign}. Results indicate that, in all of the considered scenarios, both of the Higgs bosons $H$ and $A$ are observable with signals exceeding $5\sigma$. To be specific, at the center-of-mass energy of 500 GeV, both of the Higgs bosons $H$ and $A$ are observable in the region of parameter space with $m_H=150$ GeV and $200\leq m_A \leq 250$ GeV at the integrated luminosity of 500 $fb^{-1}$. Also, at the center-of-mass energy of 1000 GeV, the $H$ Higgs boson is observable at the region with $150\leq m_H \leq 250$ GeV and $200\leq m_A \leq 330$ GeV with a mass splitting of 50-100 GeV between the $H$ and $A$ Higgs bosons at the same integrated luminosity. The $A$ Higgs boson is also observable at $\sqrt s=1000$ GeV at the same region of parameter space at an integrated luminosity up to 1000 $fb^{-1}$.
  
\section{Conclusions}  
In this study, taking the Type-\RN{1} 2HDM at the SM-like scenario as the theoretical framework and assuming several benchmark points in the parameter space of the 2HDM, observability of the predicted neutral scalar ($H$) and pseudoscalar ($A$) Higgs bosons was investigated. The process chain $e^- e^+ \rightarrow A H\rightarrow ZHH \rightarrow jj b\bar{b}b\bar{b}$ which was taken as the signal process provides an opportunity for the signal to benefit from possible enhancements Higgs-fermion couplings can receive at low $\tanb$ values. Moreover, the dependence of the $Z$-$A$-$H$ vertex upon the $\sin(\bma)$ parameter which is assumed to be unity by the SM-like assumption causes the process $e^- e^+ \rightarrow A H\rightarrow ZHH$ which contains the $Z$-$A$-$H$ vertex twice be independent of $\tanb$, and thus, the $\tanb$ value can be decreased to enhance the decay mode $H\rightarrow b\bar{b}$ without any destructive effects on the $ZHH$ production. Although the assumed hadronic decay of the $Z$ boson is expected to give rise to more errors in final results due to the uncertainties arising from jet mis-identification and mis-tag rate, the dominance of this decay channel fully compensates for the errors which arise. Several benchmark scenarios at the center-of-mass energies of 500 and 1000 GeV were assumed and event generation was performed for different scenarios separately. The beamstrahlung effect was also taken into account and the detector response was simulated based on the SiD detector at the ILC. Results show that the presented analysis can well be used to observe the scenarios considered in this study since the obtained candidate mass distributions of the Higgs bosons show significant excess of data and peaks on top of the total background near the generated masses at the assumed integrated luminosities. The integrated luminosity is set to 500 $fb^{-1}$ for all of the distributions except for the distribution of Fig. \ref{ABP5} which is obtained based on the integrated luminosity of 1000 $fb^{-1}$. Results of computation of the signal significance corresponding to the whole mass range also indicate that both of the $H$ and $A$ Higgs bosons are observable with signals exceeding $5\sigma$ in all of the considered scenarios. To be specific, the parameter space region with $m_H=150$ GeV and $200\leq m_A \leq 250$ GeV is observable at $\sqrt s=500$ GeV at the integrated luminosity of 500 $fb^{-1}$. Also, the region with $150\leq m_H \leq 250$ GeV and $200\leq m_A \leq 330$ GeV with a mass splitting of 50-100 GeV between the $H$ and $A$ Higgs bosons is observable at $\sqrt s=1000$ GeV at the same integrated luminosity. Results of the mass reconstruction which was performed by fitting proper functions to obtained mass distributions also indicate that, in all of the assumed scenarios, obtained reconstructed masses of the Higgs bosons are in reasonable agreement with generated masses and therefore, mass measurement is possible for both of the Higgs bosons. The presented analysis is expected to serve as a tool to search for the predicted 2HDM neutral Higgs bosons since not only the simulation results are promising, but also the needed center-of-mass energy and integrated luminosity for observing the considered scenarios are well below the maximum power of the future linear colliders.    
 
\section*{Acknowledgements}
We would like to thank the college of sciences at Shiraz university for providing computational facilities and maintaining the computing cluster during the research program.

%


\begin{thebibliography}{59}%
\makeatletter
\providecommand \@ifxundefined [1]{%
 \@ifx{#1\undefined}
}%
\providecommand \@ifnum [1]{%
 \ifnum #1\expandafter \@firstoftwo
 \else \expandafter \@secondoftwo
 \fi
}%
\providecommand \@ifx [1]{%
 \ifx #1\expandafter \@firstoftwo
 \else \expandafter \@secondoftwo
 \fi
}%
\providecommand \natexlab [1]{#1}%
\providecommand \enquote  [1]{``#1''}%
\providecommand \bibnamefont  [1]{#1}%
\providecommand \bibfnamefont [1]{#1}%
\providecommand \citenamefont [1]{#1}%
\providecommand \href@noop [0]{\@secondoftwo}%
\providecommand \href [0]{\begingroup \@sanitize@url \@href}%
\providecommand \@href[1]{\@@startlink{#1}\@@href}%
\providecommand \@@href[1]{\endgroup#1\@@endlink}%
\providecommand \@sanitize@url [0]{\catcode `\\12\catcode `\$12\catcode
  `\&12\catcode `\#12\catcode `\^12\catcode `\_12\catcode `\%12\relax}%
\providecommand \@@startlink[1]{}%
\providecommand \@@endlink[0]{}%
\providecommand \url  [0]{\begingroup\@sanitize@url \@url }%
\providecommand \@url [1]{\endgroup\@href {#1}{\urlprefix }}%
\providecommand \urlprefix  [0]{URL }%
\providecommand \Eprint [0]{\href }%
\providecommand \doibase [0]{http://dx.doi.org/}%
\providecommand \selectlanguage [0]{\@gobble}%
\providecommand \bibinfo  [0]{\@secondoftwo}%
\providecommand \bibfield  [0]{\@secondoftwo}%
\providecommand \translation [1]{[#1]}%
\providecommand \BibitemOpen [0]{}%
\providecommand \bibitemStop [0]{}%
\providecommand \bibitemNoStop [0]{.\EOS\space}%
\providecommand \EOS [0]{\spacefactor3000\relax}%
\providecommand \BibitemShut  [1]{\csname bibitem#1\endcsname}%
\let\auto@bib@innerbib\@empty
\bibitem [{\citenamefont {Chatrchyan}\ \emph {et~al.}(2012)\citenamefont
  {Chatrchyan} \emph {et~al.}}]{HiggsObservationCMS}%
  \BibitemOpen
  \bibfield  {author} {\bibinfo {author} {\bibfnamefont {S.}~\bibnamefont
  {Chatrchyan}} \emph {et~al.} (\bibinfo {collaboration} {CMS}),\ }\href
  {\doibase 10.1016/j.physletb.2012.08.021} {\bibfield  {journal} {\bibinfo
  {journal} {Phys. Lett.}\ }\textbf {\bibinfo {volume} {B716}},\ \bibinfo
  {pages} {30} (\bibinfo {year} {2012})},\ \Eprint
  {http://arxiv.org/abs/1207.7235} {arXiv:1207.7235 [hep-ex]} \BibitemShut
  {NoStop}%
\bibitem [{\citenamefont {Aad}\ \emph {et~al.}(2012)\citenamefont {Aad} \emph
  {et~al.}}]{HiggsObservationATLAS}%
  \BibitemOpen
  \bibfield  {author} {\bibinfo {author} {\bibfnamefont {G.}~\bibnamefont
  {Aad}} \emph {et~al.} (\bibinfo {collaboration} {ATLAS}),\ }\href {\doibase
  10.1016/j.physletb.2012.08.020} {\bibfield  {journal} {\bibinfo  {journal}
  {Phys. Lett.}\ }\textbf {\bibinfo {volume} {B716}},\ \bibinfo {pages} {1}
  (\bibinfo {year} {2012})},\ \Eprint {http://arxiv.org/abs/1207.7214}
  {arXiv:1207.7214 [hep-ex]} \BibitemShut {NoStop}%
\bibitem [{\citenamefont {Englert}\ and\ \citenamefont
  {Brout}(1964)}]{Englert1}%
  \BibitemOpen
  \bibfield  {author} {\bibinfo {author} {\bibfnamefont {F.}~\bibnamefont
  {Englert}}\ and\ \bibinfo {author} {\bibfnamefont {R.}~\bibnamefont
  {Brout}},\ }\href {\doibase 10.1103/PhysRevLett.13.321} {\bibfield  {journal}
  {\bibinfo  {journal} {Phys. Rev. Lett.}\ }\textbf {\bibinfo {volume} {13}},\
  \bibinfo {pages} {321} (\bibinfo {year} {1964})}\BibitemShut {NoStop}%
\bibitem [{\citenamefont {Higgs}(1964{\natexlab{a}})}]{Higgs1}%
  \BibitemOpen
  \bibfield  {author} {\bibinfo {author} {\bibfnamefont {P.~W.}\ \bibnamefont
  {Higgs}},\ }\href {\doibase 10.1103/PhysRevLett.13.508} {\bibfield  {journal}
  {\bibinfo  {journal} {Phys. Rev. Lett.}\ }\textbf {\bibinfo {volume} {13}},\
  \bibinfo {pages} {508} (\bibinfo {year} {1964}{\natexlab{a}})}\BibitemShut
  {NoStop}%
\bibitem [{\citenamefont {Higgs}(1964{\natexlab{b}})}]{Higgs2}%
  \BibitemOpen
  \bibfield  {author} {\bibinfo {author} {\bibfnamefont {P.~W.}\ \bibnamefont
  {Higgs}},\ }\href {\doibase 10.1016/0031-9163(64)91136-9} {\bibfield
  {journal} {\bibinfo  {journal} {Phys. Lett.}\ }\textbf {\bibinfo {volume}
  {12}},\ \bibinfo {pages} {132} (\bibinfo {year}
  {1964}{\natexlab{b}})}\BibitemShut {NoStop}%
\bibitem [{\citenamefont {Guralnik}\ \emph {et~al.}(1964)\citenamefont
  {Guralnik}, \citenamefont {Hagen},\ and\ \citenamefont {Kibble}}]{Kibble1}%
  \BibitemOpen
  \bibfield  {author} {\bibinfo {author} {\bibfnamefont {G.~S.}\ \bibnamefont
  {Guralnik}}, \bibinfo {author} {\bibfnamefont {C.~R.}\ \bibnamefont {Hagen}},
  \ and\ \bibinfo {author} {\bibfnamefont {T.~W.~B.}\ \bibnamefont {Kibble}},\
  }\href {\doibase 10.1103/PhysRevLett.13.585} {\bibfield  {journal} {\bibinfo
  {journal} {Phys. Rev. Lett.}\ }\textbf {\bibinfo {volume} {13}},\ \bibinfo
  {pages} {585} (\bibinfo {year} {1964})}\BibitemShut {NoStop}%
\bibitem [{\citenamefont {Higgs}(1966)}]{Higgs3}%
  \BibitemOpen
  \bibfield  {author} {\bibinfo {author} {\bibfnamefont {P.~W.}\ \bibnamefont
  {Higgs}},\ }\href {\doibase 10.1103/PhysRev.145.1156} {\bibfield  {journal}
  {\bibinfo  {journal} {Phys. Rev.}\ }\textbf {\bibinfo {volume} {145}},\
  \bibinfo {pages} {1156} (\bibinfo {year} {1966})}\BibitemShut {NoStop}%
\bibitem [{\citenamefont {Kibble}(1967)}]{Kibble2}%
  \BibitemOpen
  \bibfield  {author} {\bibinfo {author} {\bibfnamefont {T.~W.~B.}\
  \bibnamefont {Kibble}},\ }\href {\doibase 10.1103/PhysRev.155.1554}
  {\bibfield  {journal} {\bibinfo  {journal} {Phys. Rev.}\ }\textbf {\bibinfo
  {volume} {155}},\ \bibinfo {pages} {1554} (\bibinfo {year}
  {1967})}\BibitemShut {NoStop}%
\bibitem [{\citenamefont {Trodden}(1998)}]{Trodden}%
  \BibitemOpen
  \bibfield  {author} {\bibinfo {author} {\bibfnamefont {M.}~\bibnamefont
  {Trodden}},\ }in\ \href
  {https://inspirehep.net/record/470220/files/Pages_from_C98-03-14_471.pdf}
  {\emph {\bibinfo {booktitle} {{Proceedings, 33rd Rencontres de Moriond 98
  electrowek interactions and unified theories: Les Arcs, France, Mar 14-21,
  1998}}}}\ (\bibinfo {year} {1998})\ pp.\ \bibinfo {pages} {471--480},\
  \Eprint {http://arxiv.org/abs/hep-ph/9805252} {arXiv:hep-ph/9805252 [hep-ph]}
  \BibitemShut {NoStop}%
\bibitem [{\citenamefont {Aitchison}(2005)}]{MSSM1}%
  \BibitemOpen
  \bibfield  {author} {\bibinfo {author} {\bibfnamefont {I.~J.~R.}\
  \bibnamefont {Aitchison}},\ }\href@noop {} {\  (\bibinfo {year} {2005})},\
  \Eprint {http://arxiv.org/abs/hep-ph/0505105} {arXiv:hep-ph/0505105 [hep-ph]}
  \BibitemShut {NoStop}%
\bibitem [{\citenamefont {Kim}(1987)}]{KIM1}%
  \BibitemOpen
  \bibfield  {author} {\bibinfo {author} {\bibfnamefont {J.~E.}\ \bibnamefont
  {Kim}},\ }\href {\doibase https://doi.org/10.1016/0370-1573(87)90017-2}
  {\bibfield  {journal} {\bibinfo  {journal} {Physics Reports}\ }\textbf
  {\bibinfo {volume} {150}},\ \bibinfo {pages} {1 } (\bibinfo {year}
  {1987})}\BibitemShut {NoStop}%
\bibitem [{\citenamefont {Branco}\ \emph {et~al.}(2012)\citenamefont {Branco},
  \citenamefont {Ferreira}, \citenamefont {Lavoura}, \citenamefont {Rebelo},
  \citenamefont {Sher},\ and\ \citenamefont {Silva}}]{2hdm_TheoryPheno}%
  \BibitemOpen
  \bibfield  {author} {\bibinfo {author} {\bibfnamefont {G.~C.}\ \bibnamefont
  {Branco}}, \bibinfo {author} {\bibfnamefont {P.~M.}\ \bibnamefont
  {Ferreira}}, \bibinfo {author} {\bibfnamefont {L.}~\bibnamefont {Lavoura}},
  \bibinfo {author} {\bibfnamefont {M.~N.}\ \bibnamefont {Rebelo}}, \bibinfo
  {author} {\bibfnamefont {M.}~\bibnamefont {Sher}}, \ and\ \bibinfo {author}
  {\bibfnamefont {J.~P.}\ \bibnamefont {Silva}},\ }\href {\doibase
  10.1016/j.physrep.2012.02.002} {\bibfield  {journal} {\bibinfo  {journal}
  {Phys. Rept.}\ }\textbf {\bibinfo {volume} {516}},\ \bibinfo {pages} {1}
  (\bibinfo {year} {2012})},\ \Eprint {http://arxiv.org/abs/1106.0034}
  {arXiv:1106.0034 [hep-ph]} \BibitemShut {NoStop}%
\bibitem [{\citenamefont {Lee}(1973)}]{2hdm1}%
  \BibitemOpen
  \bibfield  {author} {\bibinfo {author} {\bibfnamefont {T.~D.}\ \bibnamefont
  {Lee}},\ }\href {\doibase 10.1103/PhysRevD.8.1226} {\bibfield  {journal}
  {\bibinfo  {journal} {Phys. Rev.}\ }\textbf {\bibinfo {volume} {D8}},\
  \bibinfo {pages} {1226} (\bibinfo {year} {1973})}\BibitemShut {NoStop}%
\bibitem [{\citenamefont {Glashow}\ and\ \citenamefont
  {Weinberg}(1977)}]{2hdm2}%
  \BibitemOpen
  \bibfield  {author} {\bibinfo {author} {\bibfnamefont {S.~L.}\ \bibnamefont
  {Glashow}}\ and\ \bibinfo {author} {\bibfnamefont {S.}~\bibnamefont
  {Weinberg}},\ }\href {\doibase 10.1103/PhysRevD.15.1958} {\bibfield
  {journal} {\bibinfo  {journal} {Phys. Rev.}\ }\textbf {\bibinfo {volume}
  {D15}},\ \bibinfo {pages} {1958} (\bibinfo {year} {1977})}\BibitemShut
  {NoStop}%
\bibitem [{\citenamefont {Branco}(1980)}]{2hdm3}%
  \BibitemOpen
  \bibfield  {author} {\bibinfo {author} {\bibfnamefont {G.~C.}\ \bibnamefont
  {Branco}},\ }\href {\doibase 10.1103/PhysRevD.22.2901} {\bibfield  {journal}
  {\bibinfo  {journal} {Phys. Rev.}\ }\textbf {\bibinfo {volume} {D22}},\
  \bibinfo {pages} {2901} (\bibinfo {year} {1980})}\BibitemShut {NoStop}%
\bibitem [{\citenamefont {Mrazek}\ \emph {et~al.}(2011)\citenamefont {Mrazek},
  \citenamefont {Pomarol}, \citenamefont {Rattazzi}, \citenamefont {Redi},
  \citenamefont {Serra},\ and\ \citenamefont {Wulzer}}]{2hdm4_CompositeHiggs}%
  \BibitemOpen
  \bibfield  {author} {\bibinfo {author} {\bibfnamefont {J.}~\bibnamefont
  {Mrazek}}, \bibinfo {author} {\bibfnamefont {A.}~\bibnamefont {Pomarol}},
  \bibinfo {author} {\bibfnamefont {R.}~\bibnamefont {Rattazzi}}, \bibinfo
  {author} {\bibfnamefont {M.}~\bibnamefont {Redi}}, \bibinfo {author}
  {\bibfnamefont {J.}~\bibnamefont {Serra}}, \ and\ \bibinfo {author}
  {\bibfnamefont {A.}~\bibnamefont {Wulzer}},\ }\href {\doibase
  10.1016/j.nuclphysb.2011.07.008} {\bibfield  {journal} {\bibinfo  {journal}
  {Nucl. Phys.}\ }\textbf {\bibinfo {volume} {B853}},\ \bibinfo {pages} {1}
  (\bibinfo {year} {2011})},\ \Eprint {http://arxiv.org/abs/1105.5403}
  {arXiv:1105.5403 [hep-ph]} \BibitemShut {NoStop}%
\bibitem [{\citenamefont {Davidson}\ and\ \citenamefont
  {Haber}(2005)}]{2hdm_HiggsSector1}%
  \BibitemOpen
  \bibfield  {author} {\bibinfo {author} {\bibfnamefont {S.}~\bibnamefont
  {Davidson}}\ and\ \bibinfo {author} {\bibfnamefont {H.~E.}\ \bibnamefont
  {Haber}},\ }\href {\doibase 10.1103/PhysRevD.72.099902,
  10.1103/PhysRevD.72.035004} {\bibfield  {journal} {\bibinfo  {journal} {Phys.
  Rev.}\ }\textbf {\bibinfo {volume} {D72}},\ \bibinfo {pages} {035004}
  (\bibinfo {year} {2005})},\ \bibinfo {note} {[Erratum: Phys.
  Rev.D72,099902(2005)]},\ \Eprint {http://arxiv.org/abs/hep-ph/0504050}
  {arXiv:hep-ph/0504050 [hep-ph]} \BibitemShut {NoStop}%
\bibitem [{\citenamefont {Aoki}\ \emph {et~al.}(2009)\citenamefont {Aoki},
  \citenamefont {Kanemura}, \citenamefont {Tsumura},\ and\ \citenamefont
  {Yagyu}}]{2hdm_HiggsSector2}%
  \BibitemOpen
  \bibfield  {author} {\bibinfo {author} {\bibfnamefont {M.}~\bibnamefont
  {Aoki}}, \bibinfo {author} {\bibfnamefont {S.}~\bibnamefont {Kanemura}},
  \bibinfo {author} {\bibfnamefont {K.}~\bibnamefont {Tsumura}}, \ and\
  \bibinfo {author} {\bibfnamefont {K.}~\bibnamefont {Yagyu}},\ }\href
  {\doibase 10.1103/PhysRevD.80.015017} {\bibfield  {journal} {\bibinfo
  {journal} {Phys. Rev.}\ }\textbf {\bibinfo {volume} {D80}},\ \bibinfo {pages}
  {015017} (\bibinfo {year} {2009})},\ \Eprint {http://arxiv.org/abs/0902.4665}
  {arXiv:0902.4665 [hep-ph]} \BibitemShut {NoStop}%
\bibitem [{\citenamefont {Campos}\ \emph {et~al.}(2017)\citenamefont {Campos},
  \citenamefont {Cogollo}, \citenamefont {Lindner}, \citenamefont {Melo},
  \citenamefont {Queiroz},\ and\ \citenamefont {Rodejohann}}]{Campos:2017dgc}%
  \BibitemOpen
  \bibfield  {author} {\bibinfo {author} {\bibfnamefont {M.~D.}\ \bibnamefont
  {Campos}}, \bibinfo {author} {\bibfnamefont {D.}~\bibnamefont {Cogollo}},
  \bibinfo {author} {\bibfnamefont {M.}~\bibnamefont {Lindner}}, \bibinfo
  {author} {\bibfnamefont {T.}~\bibnamefont {Melo}}, \bibinfo {author}
  {\bibfnamefont {F.~S.}\ \bibnamefont {Queiroz}}, \ and\ \bibinfo {author}
  {\bibfnamefont {W.}~\bibnamefont {Rodejohann}},\ }\href {\doibase
  10.1007/JHEP08(2017)092} {\bibfield  {journal} {\bibinfo  {journal} {JHEP}\
  }\textbf {\bibinfo {volume} {08}},\ \bibinfo {pages} {092} (\bibinfo {year}
  {2017})},\ \Eprint {http://arxiv.org/abs/1705.05388} {arXiv:1705.05388
  [hep-ph]} \BibitemShut {NoStop}%
\bibitem [{\citenamefont {Hashemi}\ and\ \citenamefont
  {Haghighat}(2017)}]{H-2HDMX}%
  \BibitemOpen
  \bibfield  {author} {\bibinfo {author} {\bibfnamefont {M.}~\bibnamefont
  {Hashemi}}\ and\ \bibinfo {author} {\bibfnamefont {G.}~\bibnamefont
  {Haghighat}},\ }\href {\doibase
  https://doi.org/10.1016/j.physletb.2017.06.068} {\bibfield  {journal}
  {\bibinfo  {journal} {Physics Letters B}\ }\textbf {\bibinfo {volume}
  {772}},\ \bibinfo {pages} {426 } (\bibinfo {year} {2017})}\BibitemShut
  {NoStop}%
\bibitem [{\citenamefont {Hashemi}\ and\ \citenamefont
  {Mahdavikhorrami}(2018)}]{MHashemiMMahdavi-1-4b}%
  \BibitemOpen
  \bibfield  {author} {\bibinfo {author} {\bibfnamefont {M.}~\bibnamefont
  {Hashemi}}\ and\ \bibinfo {author} {\bibfnamefont {M.}~\bibnamefont
  {Mahdavikhorrami}},\ }\href {\doibase 10.1140/epjc/s10052-018-5962-2}
  {\bibfield  {journal} {\bibinfo  {journal} {Eur. Phys. J.}\ }\textbf
  {\bibinfo {volume} {C78}},\ \bibinfo {pages} {485} (\bibinfo {year}
  {2018})},\ \Eprint {http://arxiv.org/abs/1804.10790} {arXiv:1804.10790
  [hep-ph]} \BibitemShut {NoStop}%
\bibitem [{\citenamefont {Bonvicini}\ and\ \citenamefont
  {Welch}(1998)}]{beamstrahlung}%
  \BibitemOpen
  \bibfield  {author} {\bibinfo {author} {\bibfnamefont {G.}~\bibnamefont
  {Bonvicini}}\ and\ \bibinfo {author} {\bibfnamefont {J.}~\bibnamefont
  {Welch}},\ }\href {\doibase 10.1016/S0168-9002(98)00659-7} {\bibfield
  {journal} {\bibinfo  {journal} {Nucl. Instrum. Meth.}\ }\textbf {\bibinfo
  {volume} {A418}},\ \bibinfo {pages} {223} (\bibinfo {year} {1998})},\ \Eprint
  {http://arxiv.org/abs/physics/9812023} {arXiv:physics/9812023 [physics]}
  \BibitemShut {NoStop}%
\bibitem [{\citenamefont {Potter}(2016)}]{SiDatILC}%
  \BibitemOpen
  \bibfield  {author} {\bibinfo {author} {\bibfnamefont {C.~T.}\ \bibnamefont
  {Potter}},\ }in\ \href@noop {} {\emph {\bibinfo {booktitle} {{Proceedings,
  International Workshop on Future Linear Colliders (LCWS15): Whistler, B.C.,
  Canada, November 02-06, 2015}}}}\ (\bibinfo {year} {2016})\ \Eprint
  {http://arxiv.org/abs/1602.07748} {arXiv:1602.07748 [hep-ph]} \BibitemShut
  {NoStop}%
\bibitem [{\citenamefont {Barger}\ \emph {et~al.}(1990)\citenamefont {Barger},
  \citenamefont {Hewett},\ and\ \citenamefont {Phillips}}]{Barger_2hdmTypes}%
  \BibitemOpen
  \bibfield  {author} {\bibinfo {author} {\bibfnamefont {V.~D.}\ \bibnamefont
  {Barger}}, \bibinfo {author} {\bibfnamefont {J.~L.}\ \bibnamefont {Hewett}},
  \ and\ \bibinfo {author} {\bibfnamefont {R.~J.~N.}\ \bibnamefont
  {Phillips}},\ }\href {\doibase 10.1103/PhysRevD.41.3421} {\bibfield
  {journal} {\bibinfo  {journal} {Phys. Rev.}\ }\textbf {\bibinfo {volume}
  {D41}},\ \bibinfo {pages} {3421} (\bibinfo {year} {1990})}\BibitemShut
  {NoStop}%
\bibitem [{\citenamefont {Deshpande}\ and\ \citenamefont
  {Ma}(1978)}]{Deshpande}%
  \BibitemOpen
  \bibfield  {author} {\bibinfo {author} {\bibfnamefont {N.~G.}\ \bibnamefont
  {Deshpande}}\ and\ \bibinfo {author} {\bibfnamefont {E.}~\bibnamefont {Ma}},\
  }\href {\doibase 10.1103/PhysRevD.18.2574} {\bibfield  {journal} {\bibinfo
  {journal} {Phys. Rev. D}\ }\textbf {\bibinfo {volume} {18}},\ \bibinfo
  {pages} {2574} (\bibinfo {year} {1978})}\BibitemShut {NoStop}%
\bibitem [{\citenamefont {H{\"u}ffel}\ and\ \citenamefont
  {P{\'o}csik}(1981)}]{Huffel}%
  \BibitemOpen
  \bibfield  {author} {\bibinfo {author} {\bibfnamefont {H.}~\bibnamefont
  {H{\"u}ffel}}\ and\ \bibinfo {author} {\bibfnamefont {G.}~\bibnamefont
  {P{\'o}csik}},\ }\href {\doibase 10.1007/BF01429824} {\bibfield  {journal}
  {\bibinfo  {journal} {Zeitschrift f{\"u}r Physik C Particles and Fields}\
  }\textbf {\bibinfo {volume} {8}},\ \bibinfo {pages} {13} (\bibinfo {year}
  {1981})}\BibitemShut {NoStop}%
\bibitem [{\citenamefont {Maalampi}\ \emph {et~al.}(1991)\citenamefont
  {Maalampi}, \citenamefont {Sirkka},\ and\ \citenamefont {Vilja}}]{Maalampi}%
  \BibitemOpen
  \bibfield  {author} {\bibinfo {author} {\bibfnamefont {J.}~\bibnamefont
  {Maalampi}}, \bibinfo {author} {\bibfnamefont {J.}~\bibnamefont {Sirkka}}, \
  and\ \bibinfo {author} {\bibfnamefont {I.}~\bibnamefont {Vilja}},\ }\href
  {\doibase https://doi.org/10.1016/0370-2693(91)90068-2} {\bibfield  {journal}
  {\bibinfo  {journal} {Physics Letters B}\ }\textbf {\bibinfo {volume}
  {265}},\ \bibinfo {pages} {371 } (\bibinfo {year} {1991})}\BibitemShut
  {NoStop}%
\bibitem [{\citenamefont {Kanemura}\ \emph {et~al.}(2012)\citenamefont
  {Kanemura}, \citenamefont {Tsumura},\ and\ \citenamefont
  {Yokoya}}]{KANEMURA}%
  \BibitemOpen
  \bibfield  {author} {\bibinfo {author} {\bibfnamefont {S.}~\bibnamefont
  {Kanemura}}, \bibinfo {author} {\bibfnamefont {K.}~\bibnamefont {Tsumura}}, \
  and\ \bibinfo {author} {\bibfnamefont {H.}~\bibnamefont {Yokoya}},\ }\href
  {\doibase 10.1103/PhysRevD.85.095001} {\bibfield  {journal} {\bibinfo
  {journal} {Phys. Rev.}\ }\textbf {\bibinfo {volume} {D85}},\ \bibinfo {pages}
  {095001} (\bibinfo {year} {2012})},\ \Eprint {http://arxiv.org/abs/1111.6089}
  {arXiv:1111.6089 [hep-ph]} \BibitemShut {NoStop}%
\bibitem [{\citenamefont {Akeroyd}\ \emph {et~al.}(2000)\citenamefont
  {Akeroyd}, \citenamefont {Arhrib},\ and\ \citenamefont {Naimi}}]{GAKEROYD}%
  \BibitemOpen
  \bibfield  {author} {\bibinfo {author} {\bibfnamefont {A.~G.}\ \bibnamefont
  {Akeroyd}}, \bibinfo {author} {\bibfnamefont {A.}~\bibnamefont {Arhrib}}, \
  and\ \bibinfo {author} {\bibfnamefont {E.}~\bibnamefont {Naimi}},\ }\href
  {\doibase https://doi.org/10.1016/S0370-2693(00)00962-X} {\bibfield
  {journal} {\bibinfo  {journal} {Physics Letters B}\ }\textbf {\bibinfo
  {volume} {490}},\ \bibinfo {pages} {119 } (\bibinfo {year}
  {2000})}\BibitemShut {NoStop}%
\bibitem [{\citenamefont {Eriksson}\ \emph
  {et~al.}(2010{\natexlab{a}})\citenamefont {Eriksson}, \citenamefont
  {Rathsman},\ and\ \citenamefont {Stal}}]{2hdmc1}%
  \BibitemOpen
  \bibfield  {author} {\bibinfo {author} {\bibfnamefont {D.}~\bibnamefont
  {Eriksson}}, \bibinfo {author} {\bibfnamefont {J.}~\bibnamefont {Rathsman}},
  \ and\ \bibinfo {author} {\bibfnamefont {O.}~\bibnamefont {Stal}},\ }\href
  {\doibase 10.1016/j.cpc.2009.09.011} {\bibfield  {journal} {\bibinfo
  {journal} {Comput. Phys. Commun.}\ }\textbf {\bibinfo {volume} {181}},\
  \bibinfo {pages} {189} (\bibinfo {year} {2010}{\natexlab{a}})},\ \Eprint
  {http://arxiv.org/abs/0902.0851} {arXiv:0902.0851 [hep-ph]} \BibitemShut
  {NoStop}%
\bibitem [{\citenamefont {Eriksson}\ \emph
  {et~al.}(2010{\natexlab{b}})\citenamefont {Eriksson}, \citenamefont
  {Rathsman},\ and\ \citenamefont {Stal}}]{2hdmc2}%
  \BibitemOpen
  \bibfield  {author} {\bibinfo {author} {\bibfnamefont {D.}~\bibnamefont
  {Eriksson}}, \bibinfo {author} {\bibfnamefont {J.}~\bibnamefont {Rathsman}},
  \ and\ \bibinfo {author} {\bibfnamefont {O.}~\bibnamefont {Stal}},\ }\href
  {\doibase 10.1016/j.cpc.2009.12.016} {\bibfield  {journal} {\bibinfo
  {journal} {Comput. Phys. Commun.}\ }\textbf {\bibinfo {volume} {181}},\
  \bibinfo {pages} {833} (\bibinfo {year} {2010}{\natexlab{b}})}\BibitemShut
  {NoStop}%
\bibitem [{\citenamefont {Bechtle}\ \emph
  {et~al.}(2014{\natexlab{a}})\citenamefont {Bechtle}, \citenamefont {Brein},
  \citenamefont {Heinemeyer}, \citenamefont {Stål}, \citenamefont {Stefaniak},
  \citenamefont {Weiglein},\ and\ \citenamefont {Williams}}]{HiggsBounds4.3.1}%
  \BibitemOpen
  \bibfield  {author} {\bibinfo {author} {\bibfnamefont {P.}~\bibnamefont
  {Bechtle}}, \bibinfo {author} {\bibfnamefont {O.}~\bibnamefont {Brein}},
  \bibinfo {author} {\bibfnamefont {S.}~\bibnamefont {Heinemeyer}}, \bibinfo
  {author} {\bibfnamefont {O.}~\bibnamefont {Stål}}, \bibinfo {author}
  {\bibfnamefont {T.}~\bibnamefont {Stefaniak}}, \bibinfo {author}
  {\bibfnamefont {G.}~\bibnamefont {Weiglein}}, \ and\ \bibinfo {author}
  {\bibfnamefont {K.~E.}\ \bibnamefont {Williams}},\ }\href {\doibase
  10.1140/epjc/s10052-013-2693-2} {\bibfield  {journal} {\bibinfo  {journal}
  {Eur. Phys. J.}\ }\textbf {\bibinfo {volume} {C74}},\ \bibinfo {pages} {2693}
  (\bibinfo {year} {2014}{\natexlab{a}})},\ \Eprint
  {http://arxiv.org/abs/1311.0055} {arXiv:1311.0055 [hep-ph]} \BibitemShut
  {NoStop}%
\bibitem [{\citenamefont {Bechtle}\ \emph
  {et~al.}(2014{\natexlab{b}})\citenamefont {Bechtle}, \citenamefont
  {Heinemeyer}, \citenamefont {Stål}, \citenamefont {Stefaniak},\ and\
  \citenamefont {Weiglein}}]{HiggsSignals1.3.0}%
  \BibitemOpen
  \bibfield  {author} {\bibinfo {author} {\bibfnamefont {P.}~\bibnamefont
  {Bechtle}}, \bibinfo {author} {\bibfnamefont {S.}~\bibnamefont {Heinemeyer}},
  \bibinfo {author} {\bibfnamefont {O.}~\bibnamefont {Stål}}, \bibinfo
  {author} {\bibfnamefont {T.}~\bibnamefont {Stefaniak}}, \ and\ \bibinfo
  {author} {\bibfnamefont {G.}~\bibnamefont {Weiglein}},\ }\href {\doibase
  10.1140/epjc/s10052-013-2711-4} {\bibfield  {journal} {\bibinfo  {journal}
  {Eur. Phys. J.}\ }\textbf {\bibinfo {volume} {C74}},\ \bibinfo {pages} {2711}
  (\bibinfo {year} {2014}{\natexlab{b}})},\ \Eprint
  {http://arxiv.org/abs/1305.1933} {arXiv:1305.1933 [hep-ph]} \BibitemShut
  {NoStop}%
\bibitem [{\citenamefont {Bertolini}(1986)}]{BERTOLINI}%
  \BibitemOpen
  \bibfield  {author} {\bibinfo {author} {\bibfnamefont {S.}~\bibnamefont
  {Bertolini}},\ }\href {\doibase https://doi.org/10.1016/0550-3213(86)90341-X}
  {\bibfield  {journal} {\bibinfo  {journal} {Nuclear Physics B}\ }\textbf
  {\bibinfo {volume} {272}},\ \bibinfo {pages} {77 } (\bibinfo {year}
  {1986})}\BibitemShut {NoStop}%
\bibitem [{\citenamefont {Denner}\ \emph {et~al.}(1990)\citenamefont {Denner},
  \citenamefont {Guth},\ and\ \citenamefont {Kühn}}]{DENNER}%
  \BibitemOpen
  \bibfield  {author} {\bibinfo {author} {\bibfnamefont {A.}~\bibnamefont
  {Denner}}, \bibinfo {author} {\bibfnamefont {R.}~\bibnamefont {Guth}}, \ and\
  \bibinfo {author} {\bibfnamefont {J.}~\bibnamefont {Kühn}},\ }\href
  {\doibase https://doi.org/10.1016/0370-2693(90)91126-V} {\bibfield  {journal}
  {\bibinfo  {journal} {Physics Letters B}\ }\textbf {\bibinfo {volume}
  {240}},\ \bibinfo {pages} {438 } (\bibinfo {year} {1990})}\BibitemShut
  {NoStop}%
\bibitem [{\citenamefont {et~al}(2006)}]{Yao}%
  \BibitemOpen
  \bibfield  {author} {\bibinfo {author} {\bibfnamefont {W.-M.~Y.}\
  \bibnamefont {et~al}},\ }\href {http://stacks.iop.org/0954-3899/33/i=1/a=001}
  {\bibfield  {journal} {\bibinfo  {journal} {Journal of Physics G: Nuclear and
  Particle Physics}\ }\textbf {\bibinfo {volume} {33}},\ \bibinfo {pages} {1}
  (\bibinfo {year} {2006})}\BibitemShut {NoStop}%
\bibitem [{\citenamefont {Grimus}\ \emph {et~al.}(2008)\citenamefont {Grimus},
  \citenamefont {Lavoura}, \citenamefont {Ogreid},\ and\ \citenamefont
  {Osland}}]{drho}%
  \BibitemOpen
  \bibfield  {author} {\bibinfo {author} {\bibfnamefont {W.}~\bibnamefont
  {Grimus}}, \bibinfo {author} {\bibfnamefont {L.}~\bibnamefont {Lavoura}},
  \bibinfo {author} {\bibfnamefont {O.~M.}\ \bibnamefont {Ogreid}}, \ and\
  \bibinfo {author} {\bibfnamefont {P.}~\bibnamefont {Osland}},\ }\href
  {\doibase 10.1088/0954-3899/35/7/075001} {\bibfield  {journal} {\bibinfo
  {journal} {J. Phys.}\ }\textbf {\bibinfo {volume} {G35}},\ \bibinfo {pages}
  {075001} (\bibinfo {year} {2008})},\ \Eprint {http://arxiv.org/abs/0711.4022}
  {arXiv:0711.4022 [hep-ph]} \BibitemShut {NoStop}%
\bibitem [{\citenamefont {Gerard}\ and\ \citenamefont
  {Herquet}(2007)}]{Gerard:2007kn}%
  \BibitemOpen
  \bibfield  {author} {\bibinfo {author} {\bibfnamefont {J.~M.}\ \bibnamefont
  {Gerard}}\ and\ \bibinfo {author} {\bibfnamefont {M.}~\bibnamefont
  {Herquet}},\ }\href {\doibase 10.1103/PhysRevLett.98.251802} {\bibfield
  {journal} {\bibinfo  {journal} {Phys. Rev. Lett.}\ }\textbf {\bibinfo
  {volume} {98}},\ \bibinfo {pages} {251802} (\bibinfo {year} {2007})},\
  \Eprint {http://arxiv.org/abs/hep-ph/0703051} {arXiv:hep-ph/0703051 [HEP-PH]}
  \BibitemShut {NoStop}%
\bibitem [{\citenamefont {Aaboud}\ \emph {et~al.}(2018)\citenamefont {Aaboud}
  \emph {et~al.}}]{ATLAS-2HDM-2}%
  \BibitemOpen
  \bibfield  {author} {\bibinfo {author} {\bibfnamefont {M.}~\bibnamefont
  {Aaboud}} \emph {et~al.} (\bibinfo {collaboration} {ATLAS}),\ }\href@noop {}
  {\  (\bibinfo {year} {2018})},\ \Eprint {http://arxiv.org/abs/1804.01126}
  {arXiv:1804.01126 [hep-ex]} \BibitemShut {NoStop}%
\bibitem [{CMS(2018)}]{CMS-2HDM-2}%
  \BibitemOpen
  \href {http://cds.cern.ch/record/2628545} {\emph {\bibinfo {title} {{Search
  for a heavy pseudoscalar boson decaying to a Z boson and a Higgs boson at
  sqrt(s)=13 TeV}}}},\ \bibinfo {type} {Tech. Rep.}\ \bibinfo {number}
  {CMS-PAS-HIG-18-005}\ (\bibinfo  {institution} {CERN},\ \bibinfo {address}
  {Geneva},\ \bibinfo {year} {2018})\BibitemShut {NoStop}%
\bibitem [{\citenamefont {Aad}\ \emph {et~al.}(2015)\citenamefont {Aad} \emph
  {et~al.}}]{ATLAS-2HDM-1}%
  \BibitemOpen
  \bibfield  {author} {\bibinfo {author} {\bibfnamefont {G.}~\bibnamefont
  {Aad}} \emph {et~al.} (\bibinfo {collaboration} {ATLAS}),\ }\href {\doibase
  10.1016/j.physletb.2015.03.054} {\bibfield  {journal} {\bibinfo  {journal}
  {Phys. Lett.}\ }\textbf {\bibinfo {volume} {B744}},\ \bibinfo {pages} {163}
  (\bibinfo {year} {2015})},\ \Eprint {http://arxiv.org/abs/1502.04478}
  {arXiv:1502.04478 [hep-ex]} \BibitemShut {NoStop}%
\bibitem [{\citenamefont {Aad}\ \emph {et~al.}(2016)\citenamefont {Aad} \emph
  {et~al.}}]{ATLAS-2HDM-3}%
  \BibitemOpen
  \bibfield  {author} {\bibinfo {author} {\bibfnamefont {G.}~\bibnamefont
  {Aad}} \emph {et~al.} (\bibinfo {collaboration} {ATLAS}),\ }\href {\doibase
  10.1140/epjc/s10052-015-3820-z} {\bibfield  {journal} {\bibinfo  {journal}
  {Eur. Phys. J.}\ }\textbf {\bibinfo {volume} {C76}},\ \bibinfo {pages} {45}
  (\bibinfo {year} {2016})},\ \Eprint {http://arxiv.org/abs/1507.05930}
  {arXiv:1507.05930 [hep-ex]} \BibitemShut {NoStop}%
\bibitem [{\citenamefont {Misiak}\ \emph {et~al.}(2015)\citenamefont {Misiak}
  \emph {et~al.}}]{Misiak}%
  \BibitemOpen
  \bibfield  {author} {\bibinfo {author} {\bibfnamefont {M.}~\bibnamefont
  {Misiak}} \emph {et~al.},\ }\href {\doibase 10.1103/PhysRevLett.114.221801}
  {\bibfield  {journal} {\bibinfo  {journal} {Phys. Rev. Lett.}\ }\textbf
  {\bibinfo {volume} {114}},\ \bibinfo {pages} {221801} (\bibinfo {year}
  {2015})},\ \Eprint {http://arxiv.org/abs/1503.01789} {arXiv:1503.01789
  [hep-ph]} \BibitemShut {NoStop}%
\bibitem [{\citenamefont {Misiak}\ and\ \citenamefont
  {Steinhauser}(2017)}]{Misiak:2017bgg}%
  \BibitemOpen
  \bibfield  {author} {\bibinfo {author} {\bibfnamefont {M.}~\bibnamefont
  {Misiak}}\ and\ \bibinfo {author} {\bibfnamefont {M.}~\bibnamefont
  {Steinhauser}},\ }\href {\doibase 10.1140/epjc/s10052-017-4776-y} {\bibfield
  {journal} {\bibinfo  {journal} {Eur. Phys. J.}\ }\textbf {\bibinfo {volume}
  {C77}},\ \bibinfo {pages} {201} (\bibinfo {year} {2017})},\ \Eprint
  {http://arxiv.org/abs/1702.04571} {arXiv:1702.04571 [hep-ph]} \BibitemShut
  {NoStop}%
\bibitem [{\citenamefont {Moretti}(2016)}]{Moretti:2016qcc}%
  \BibitemOpen
  \bibfield  {author} {\bibinfo {author} {\bibfnamefont {S.}~\bibnamefont
  {Moretti}},\ }\bibfield  {booktitle} {\emph {\bibinfo {booktitle}
  {{Proceedings, 6th International Workshop on Prospects for Charged Higgs
  Discovery at Colliders (CHARGED 2016): Uppsala, Sweden, October 3-6,
  2016}}},\ }\href@noop {} {\bibfield  {journal} {\bibinfo  {journal} {PoS}\
  }\textbf {\bibinfo {volume} {CHARGED2016}},\ \bibinfo {pages} {014} (\bibinfo
  {year} {2016})},\ \Eprint {http://arxiv.org/abs/1612.02063} {arXiv:1612.02063
  [hep-ph]} \BibitemShut {NoStop}%
\bibitem [{\citenamefont {Barate}\ \emph {et~al.}(2000)\citenamefont {Barate}
  \emph {et~al.}}]{lep1}%
  \BibitemOpen
  \bibfield  {author} {\bibinfo {author} {\bibfnamefont {R.}~\bibnamefont
  {Barate}} \emph {et~al.} (\bibinfo {collaboration} {ALEPH}),\ }\href
  {\doibase 10.1016/S0370-2693(00)00822-4} {\bibfield  {journal} {\bibinfo
  {journal} {Phys. Lett.}\ }\textbf {\bibinfo {volume} {B487}},\ \bibinfo
  {pages} {253} (\bibinfo {year} {2000})},\ \Eprint
  {http://arxiv.org/abs/hep-ex/0008005} {arXiv:hep-ex/0008005 [hep-ex]}
  \BibitemShut {NoStop}%
\bibitem [{\citenamefont {Acciarri}\ \emph {et~al.}(2000)\citenamefont
  {Acciarri} \emph {et~al.}}]{lep2}%
  \BibitemOpen
  \bibfield  {author} {\bibinfo {author} {\bibfnamefont {M.}~\bibnamefont
  {Acciarri}} \emph {et~al.} (\bibinfo {collaboration} {L3}),\ }\href {\doibase
  10.1016/S0370-2693(00)01272-7} {\bibfield  {journal} {\bibinfo  {journal}
  {Phys. Lett.}\ }\textbf {\bibinfo {volume} {B496}},\ \bibinfo {pages} {34}
  (\bibinfo {year} {2000})},\ \Eprint {http://arxiv.org/abs/hep-ex/0009010}
  {arXiv:hep-ex/0009010 [hep-ex]} \BibitemShut {NoStop}%
\bibitem [{lep()}]{lepexclusion2}%
  \BibitemOpen
  in\ \href@noop {} {\emph {\bibinfo {booktitle} {{Lepton and photon
  interactions at high energies. Proceedings, 20th International Symposium, LP
  2001, Rome, Italy, July 23-28, 2001}}}}\BibitemShut {NoStop}%
\bibitem [{CMS()}]{CMSNeutralHiggs}%
  \BibitemOpen
  \href@noop {} {\bibinfo  {journal} {CMS Collaboration, CMS-PAS-HIG-16-037}\
  }\BibitemShut {NoStop}%
\bibitem [{ATL()}]{ATLASNeutralHiggs}%
  \BibitemOpen
\bibfield  {journal} {  }\href@noop {} {\bibinfo  {journal} {The ATLAS
  Collaboration, ATLAS-CONF-2016-085}\ }\BibitemShut {NoStop}%
\bibitem [{\citenamefont {Boos}\ \emph {et~al.}(2004)\citenamefont {Boos},
  \citenamefont {Bunichev}, \citenamefont {Dubinin}, \citenamefont {Dudko},
  \citenamefont {Ilyin}, \citenamefont {Kryukov}, \citenamefont {Edneral},
  \citenamefont {Savrin}, \citenamefont {Semenov},\ and\ \citenamefont
  {Sherstnev}}]{CompHEP4.4}%
  \BibitemOpen
\bibfield  {journal} {  }\bibfield  {author} {\bibinfo {author} {\bibfnamefont
  {E.}~\bibnamefont {Boos}}, \bibinfo {author} {\bibfnamefont {V.}~\bibnamefont
  {Bunichev}}, \bibinfo {author} {\bibfnamefont {M.}~\bibnamefont {Dubinin}},
  \bibinfo {author} {\bibfnamefont {L.}~\bibnamefont {Dudko}}, \bibinfo
  {author} {\bibfnamefont {V.}~\bibnamefont {Ilyin}}, \bibinfo {author}
  {\bibfnamefont {A.}~\bibnamefont {Kryukov}}, \bibinfo {author} {\bibfnamefont
  {V.}~\bibnamefont {Edneral}}, \bibinfo {author} {\bibfnamefont
  {V.}~\bibnamefont {Savrin}}, \bibinfo {author} {\bibfnamefont
  {A.}~\bibnamefont {Semenov}}, \ and\ \bibinfo {author} {\bibfnamefont
  {A.}~\bibnamefont {Sherstnev}} (\bibinfo {collaboration} {CompHEP}),\
  }\bibfield  {booktitle} {\emph {\bibinfo {booktitle} {{Advanced computing and
  analysis techniques in physics research. Proceedings, 9th International
  Workshop, ACAT'03, Tsukuba, Japan, December 1-5, 2003}}},\ }\href {\doibase
  10.1016/j.nima.2004.07.096} {\bibfield  {journal} {\bibinfo  {journal} {Nucl.
  Instrum. Meth.}\ }\textbf {\bibinfo {volume} {A534}},\ \bibinfo {pages} {250}
  (\bibinfo {year} {2004})},\ \Eprint {http://arxiv.org/abs/hep-ph/0403113}
  {arXiv:hep-ph/0403113 [hep-ph]} \BibitemShut {NoStop}%
\bibitem [{\citenamefont {Adolphsen}\ \emph {et~al.}(2013)\citenamefont
  {Adolphsen} \emph {et~al.}}]{ILCtdrv3.II}%
  \BibitemOpen
  \bibfield  {author} {\bibinfo {author} {\bibfnamefont {C.}~\bibnamefont
  {Adolphsen}} \emph {et~al.},\ }\href {https://cds.cern.ch/record/1601969}
  {\emph {\bibinfo {title} {{The International Linear Collider Technical Design
  Report - Volume 3.II: Accelerator Baseline Design}}}},\ \bibinfo {type}
  {Tech. Rep.}\ \bibinfo {number} {arXiv:1306.6328}\ (\bibinfo {address}
  {Geneva},\ \bibinfo {year} {2013})\ \bibinfo {note} {comments: See also
  http://www.linearcollider.org/ILC/TDR . The full list of signatories is
  inside the Report}\BibitemShut {NoStop}%
\bibitem [{\citenamefont {Sjöstrand}\ \emph {et~al.}(2015)\citenamefont
  {Sjöstrand}, \citenamefont {Ask}, \citenamefont {Christiansen},
  \citenamefont {Corke}, \citenamefont {Desai}, \citenamefont {Ilten},
  \citenamefont {Mrenna}, \citenamefont {Prestel}, \citenamefont {Rasmussen},\
  and\ \citenamefont {Skands}}]{pythia82}%
  \BibitemOpen
  \bibfield  {author} {\bibinfo {author} {\bibfnamefont {T.}~\bibnamefont
  {Sjöstrand}}, \bibinfo {author} {\bibfnamefont {S.}~\bibnamefont {Ask}},
  \bibinfo {author} {\bibfnamefont {J.~R.}\ \bibnamefont {Christiansen}},
  \bibinfo {author} {\bibfnamefont {R.}~\bibnamefont {Corke}}, \bibinfo
  {author} {\bibfnamefont {N.}~\bibnamefont {Desai}}, \bibinfo {author}
  {\bibfnamefont {P.}~\bibnamefont {Ilten}}, \bibinfo {author} {\bibfnamefont
  {S.}~\bibnamefont {Mrenna}}, \bibinfo {author} {\bibfnamefont
  {S.}~\bibnamefont {Prestel}}, \bibinfo {author} {\bibfnamefont {C.~O.}\
  \bibnamefont {Rasmussen}}, \ and\ \bibinfo {author} {\bibfnamefont {P.~Z.}\
  \bibnamefont {Skands}},\ }\href {\doibase
  https://doi.org/10.1016/j.cpc.2015.01.024} {\bibfield  {journal} {\bibinfo
  {journal} {Computer Physics Communications}\ }\textbf {\bibinfo {volume}
  {191}},\ \bibinfo {pages} {159 } (\bibinfo {year} {2015})}\BibitemShut
  {NoStop}%
\bibitem [{\citenamefont {de~Favereau}\ \emph {et~al.}(2014)\citenamefont
  {de~Favereau}, \citenamefont {Delaere}, \citenamefont {Demin}, \citenamefont
  {Giammanco}, \citenamefont {Lemaître}, \citenamefont {Mertens},\ and\
  \citenamefont {Selvaggi}}]{DELPHES3.4}%
  \BibitemOpen
  \bibfield  {author} {\bibinfo {author} {\bibfnamefont {J.}~\bibnamefont
  {de~Favereau}}, \bibinfo {author} {\bibfnamefont {C.}~\bibnamefont
  {Delaere}}, \bibinfo {author} {\bibfnamefont {P.}~\bibnamefont {Demin}},
  \bibinfo {author} {\bibfnamefont {A.}~\bibnamefont {Giammanco}}, \bibinfo
  {author} {\bibfnamefont {V.}~\bibnamefont {Lemaître}}, \bibinfo {author}
  {\bibfnamefont {A.}~\bibnamefont {Mertens}}, \ and\ \bibinfo {author}
  {\bibfnamefont {M.}~\bibnamefont {Selvaggi}} (\bibinfo {collaboration}
  {DELPHES 3}),\ }\href {\doibase 10.1007/JHEP02(2014)057} {\bibfield
  {journal} {\bibinfo  {journal} {JHEP}\ }\textbf {\bibinfo {volume} {02}},\
  \bibinfo {pages} {057} (\bibinfo {year} {2014})},\ \Eprint
  {http://arxiv.org/abs/1307.6346} {arXiv:1307.6346 [hep-ex]} \BibitemShut
  {NoStop}%
\bibitem [{\citenamefont {Cacciari}\ \emph {et~al.}(2008)\citenamefont
  {Cacciari}, \citenamefont {Salam},\ and\ \citenamefont {Soyez}}]{antikt}%
  \BibitemOpen
  \bibfield  {author} {\bibinfo {author} {\bibfnamefont {M.}~\bibnamefont
  {Cacciari}}, \bibinfo {author} {\bibfnamefont {G.~P.}\ \bibnamefont {Salam}},
  \ and\ \bibinfo {author} {\bibfnamefont {G.}~\bibnamefont {Soyez}},\ }\href
  {\doibase 10.1088/1126-6708/2008/04/063} {\bibfield  {journal} {\bibinfo
  {journal} {JHEP}\ }\textbf {\bibinfo {volume} {04}},\ \bibinfo {pages} {063}
  (\bibinfo {year} {2008})},\ \Eprint {http://arxiv.org/abs/0802.1189}
  {arXiv:0802.1189 [hep-ph]} \BibitemShut {NoStop}%
\bibitem [{\citenamefont {Cacciari}(2006)}]{fastjet1}%
  \BibitemOpen
  \bibfield  {author} {\bibinfo {author} {\bibfnamefont {M.}~\bibnamefont
  {Cacciari}},\ }in\ \href@noop {} {\emph {\bibinfo {booktitle} {{Deep
  inelastic scattering. Proceedings, 14th International Workshop, DIS 2006,
  Tsukuba, Japan, April 20-24, 2006}}}}\ (\bibinfo {year} {2006})\ pp.\
  \bibinfo {pages} {487--490},\ \bibinfo {note} {[,125(2006)]},\ \Eprint
  {http://arxiv.org/abs/hep-ph/0607071} {arXiv:hep-ph/0607071 [hep-ph]}
  \BibitemShut {NoStop}%
\bibitem [{\citenamefont {Cacciari}\ \emph {et~al.}(2012)\citenamefont
  {Cacciari}, \citenamefont {Salam},\ and\ \citenamefont {Soyez}}]{fastjet2}%
  \BibitemOpen
  \bibfield  {author} {\bibinfo {author} {\bibfnamefont {M.}~\bibnamefont
  {Cacciari}}, \bibinfo {author} {\bibfnamefont {G.~P.}\ \bibnamefont {Salam}},
  \ and\ \bibinfo {author} {\bibfnamefont {G.}~\bibnamefont {Soyez}},\ }\href
  {\doibase 10.1140/epjc/s10052-012-1896-2} {\bibfield  {journal} {\bibinfo
  {journal} {Eur. Phys. J.}\ }\textbf {\bibinfo {volume} {C72}},\ \bibinfo
  {pages} {1896} (\bibinfo {year} {2012})},\ \Eprint
  {http://arxiv.org/abs/1111.6097} {arXiv:1111.6097 [hep-ph]} \BibitemShut
  {NoStop}%
\bibitem [{\citenamefont {Werlen}\ and\ \citenamefont
  {Perret-Gallix}(1997)}]{root1}%
  \BibitemOpen
  \bibfield  {author} {\bibinfo {author} {\bibfnamefont {M.}~\bibnamefont
  {Werlen}}\ and\ \bibinfo {author} {\bibfnamefont {D.}~\bibnamefont
  {Perret-Gallix}},\ }\href@noop {} {\bibfield  {journal} {\bibinfo  {journal}
  {Nucl. Instrum. Meth.}\ }\textbf {\bibinfo {volume} {A389}},\ \bibinfo
  {pages} {pp.1} (\bibinfo {year} {1997})}\BibitemShut {NoStop}%
\bibitem [{\citenamefont {Brun}\ and\ \citenamefont
  {Rademakers}(1997)}]{root2}%
  \BibitemOpen
  \bibfield  {author} {\bibinfo {author} {\bibfnamefont {R.}~\bibnamefont
  {Brun}}\ and\ \bibinfo {author} {\bibfnamefont {F.}~\bibnamefont
  {Rademakers}},\ }\bibfield  {booktitle} {\emph {\bibinfo {booktitle} {{New
  computing techniques in physics research V. Proceedings, 5th International
  Workshop, AIHENP '96, Lausanne, Switzerland, September 2-6, 1996}}},\ }\href
  {\doibase 10.1016/S0168-9002(97)00048-X} {\bibfield  {journal} {\bibinfo
  {journal} {Nucl. Instrum. Meth.}\ }\textbf {\bibinfo {volume} {A389}},\
  \bibinfo {pages} {81} (\bibinfo {year} {1997})}\BibitemShut {NoStop}%
\end{thebibliography}


\end{document}